\documentclass[useAMS,usenatbib]{mn2e}

\usepackage{aas_macros}

\usepackage{amsmath}
\usepackage{amssymb}
\usepackage{color}
\usepackage{array}
\usepackage{hhline}
\usepackage{graphicx}
\usepackage{float}

\usepackage{booktabs}
\newcommand{\ra}[1]{\renewcommand{\arraystretch}{#1}}

\usepackage{ulem}
\definecolor{darkblue}{rgb}{0,0,0.5}
\definecolor{darkgreen}{rgb}{0,0.5,0}
\definecolor{darkred}{rgb}{0.5,0,0}
\newcommand{\CRKadd}[1]{#1}
\begin{document}

\title[Uncertainties in pixel-based source reconstruction]{Statistical and systematic uncertainties in pixel-based source reconstruction algorithms for gravitational lensing}
\author[Tagore \& Keeton]{Amitpal S.\ Tagore and Charles R.\ Keeton \\
Department of Physics \& Astronomy, Rutgers University, 136 Frelinghuysen Road, Piscataway, NJ 08854, USA}
\maketitle

\begin{abstract}
Gravitational lens modeling of spatially resolved sources is a challenging inverse problem with many observational constraints and model parameters.
We examine established pixel-based source reconstruction algorithms for de-lensing the source and constraining lens model parameters.
Using test data for four canonical lens configurations, we explore statistical and systematic uncertainties associated with gridding, source regularisation, interpolation errors, noise, and telescope pointing.
Specifically, we compare two gridding schemes in the source plane: a fully adaptive grid that follows the lens mapping but is irregular, and an adaptive Cartesian grid.
We also consider regularisation schemes that minimise derivatives of the source (using two finite difference methods) and introduce a scheme that minimises deviations from an analytic source profile.
Careful choice of gridding and regularisation can reduce ``discreteness noise'' in the $\chi^2$ surface that is inherent in the pixel-based methodology.
With a gridded source, some degree of interpolation is unavoidable, and errors due to interpolation need to be taken into account (especially for high signal-to-noise data).
Different realisations of the noise and telescope pointing lead to slightly different values for lens model parameters, and the scatter between different ``observations'' can be comparable to or larger than the model uncertainties themselves.
The same effects create scatter in the lensing magnification at the level of a few percent for a peak signal-to-noise ratio of 10, which decreases as the data quality improves.

\bigskip
\end{abstract}


%%%%%%%%%%%%%%%%%%%%%%%%%%%%%%%%%%%%%%%%%%%%%%%%%%%%%%%%%%%%
\section{Introduction}
\label{sec:intro}
%%%%%%%%%%%%%%%%%%%%%%%%%%%%%%%%%%%%%%%%%%%%%%%%%%%%%%%%%%%%

The gravitational deflection of light produces a variety of observable effects that can be used to study the mass distributions of deflectors (e.g., galaxies and clusters of galaxies) and the structure of light sources (e.g., distant quasars and star-forming galaxies), and to constrain cosmological parameters \citep[see the review by][]{saasfee}.
In this paper, we focus on strong gravitational lensing in which light bending creates multiple images of the source.

If the source is compact and unresolved, the images and source are each characterised by just three numbers: two position coordinates and a flux.
If the source is extended, the resolved images provide many constraints but the structure of the source must be included in the modeling.
One approach is to assume the source has elliptical symmetry and analyse isophotal shapes \citep[e.g.,][]{blandfordEring} or peak surface brightness curves \citep[e.g.,][]{kochanekEring} in Einstein rings.
A more general approach is to reconstruct the source on a grid in order to permit complex structure and reproduce the data pixel by pixel.
Pixel-based source reconstruction (PBSR) algorithms take full advantage of the information in the lensed images, but the large numbers of constraints (image pixels) and free parameters (source pixels) demand advanced techniques and more computational effort.

The history of extended image lens modeling is rich.
Early implementations of PBSR algorithms \citep[e.g.,][]{wallington1996,koopmans:cdmsubstructure} used a two-loop method in which an outer loop varied the lens model parameters, while an inner loop varied source parameters to find the best fit given a lens model.
The lens was described parametrically, typically using standard galaxy and halo mass profiles.
The source, by contrast, was constructed on a Cartesian grid, and a penalty function was used to disfavor source models that seemed too unphysical.
Varying all of the source pixels independently was a costly step.
\citet{citewarrendyesemilinear} simplified the inner loop by showing that the lensing equation can be written as a matrix equation, allowing the optimal source to be found in a single, analytic step (see \S\ref{sec:bayesian}).
To improve the spatial resolution, \citet{dyegrid} and \citet{vegettigrid} introduced irregular source grids while keeping the inner loop linear (see \S\ref{gridsection}).

As the number of approaches to lens modeling grew, \citet{citebrewer2006} used a Bayesian framework to argue that the methods are basically equivalent and differ only in the choice of priors.
\citet{suyureg} extended the framework, further developing the idea of using a penalty function to ``regularise'' the source, and determining the strength of regularisation using Bayesian inference.
Both \citet{citebrewer2006} and \citet{suyureg} showed that the choice of prior depends on the data and the unlensed source.

To date, there have been many applications of PBSR for both lens-plane and source-plane science.
\citet{suyu:B1608a,suyu:B1608b} simultaneously reconstruct the mass distribution of the lens B1608+656 and combine the lens model with the measured time delays to constrain the Hubble constant.
\citet{suyu:B1933} disentangle the disk, bulge, and halo components in the lens B1933+503.
\citet{suyu:substr}, \citet{vegetticlone}, and \citet{vegetti:dark, vegetti:dark2} all show that mass substructure in lenses can be detected through its effects on lensed images.
\citet{sourceplanescience} and \citet{sourceplanescience2} explore the intrinsic properties of lensed high-redshift sources galaxies.

We note that additional techniques have been developed for analysing radio observations of lensed systems.
Because radio interferometers sample the visibility function (the Fourier transform of the sky brightness), radio astronomy has put much effort into developing reliable reduction algorithms.
The \textsc{CLEAN} algorithm \citep{clean} fits the ``dirty'' map of observed surface brightnesses with point sources.
It finds the brightest region in the map and subtracts a point source convolved with the instrumental beam, and then iterates until a stopping criterion is met.
LensClean \citep{lensclean} adds a step in which the point source is gravitationally lensed before the images are subtracted, allowing the lens model and source to be fit simultaneously.

In this paper, we present a new software called \textit{pixsrc} that performs PBSR in conjunction with the established \textit{lensmodel} software \citep{citelensmodel} for exploring the lens model parameter space.
We present the methodology behind \textit{pixsrc} and then discuss issues that arise during the lens modeling process.
In particular, we investigate statistical uncertainties and systematic biases inherent in PBSR methods by analysing representative galaxy-galaxy strong lensing events.
We examine the effects of noise and telescope pointing on the lens model analysis, as well the effects of different choices of gridding and priors.

%%%%%%%%%%%%%%%%%%%%%%%%%%%%%%%%%%%%%%%%%%%%%%%%%%%%%%%%%%%%
\section{Bayesian framework}
\label{sec:bayesian}
%%%%%%%%%%%%%%%%%%%%%%%%%%%%%%%%%%%%%%%%%%%%%%%%%%%%%%%%%%%%

For a given dataset, there may be lens models that fit the data well but require a source that seems unrealistic.
(A model with no mass can fit the data perfectly if the source looks exactly like the image.)
There may also be models for which the source fits the noise in addition to the lens data.
Using Bayesian inference, priors can be used to reject models that are unphysical or overfit the noise.
This section reviews the formal framework for PBSR, which has been discussed in detail by \citet{suyureg} and \citet{vegettigrid}.
We reproduce only the key aspects here.

%%%%%%%%%%%%%%%%%%%%%%%%%%%%%%%%%%%%%%%%%%%%%%%%%%%%%%%%%%%%
\subsection{Most likely solution}

In the absence of dust or other attenuation, lensing conserves surface brightness.
The mapping between the source plane and image plane can therefore be written as\footnote{We adopt the following conventions: one-dimensional vectors are denoted by bold lower-case letters, two-dimensional matrices are denoted by bold capital letters, and scalars are unbolded.}
\begin{equation}
\label{lensequation1}
\mathbf{d}=\mathbf{L}\mathbf{s} + \mathbf{n},
\end{equation}
where $\mathbf{L}$ is a linear ``lensing operator'' that acts on surface brightness values.
This operator can encode not only the gravitational deflections of the lens but also effects from the atmosphere and telescope.
For example, if $\mathbf{G}$ characterises the lens while $\mathbf{B}$ is a ``blurring operator'' that characterises the point spread function (PSF) of the observations, we can define $\mathbf{L} \equiv \mathbf{BG}$ to capture the combined effects.
$\mathbf{s}$ and $\mathbf{d}$ are vectors containing the surface brightness values in the source plane and image plane, respectively, and $\mathbf{n}$ is the noise present in the data.
If the source and data are two-dimensional images with surface brightness values specified on a Cartesian grid, the one-dimensional vectors $\mathbf{s}$ and $\mathbf{d}$ can be constructed by column- or row-stacking the two-dimensional images.
If the source grid is irregular, the structure of $\mathbf{s}$ can be more complicated, but the formal framework still applies.
For reference, we note that the numbers of pixels in the source and image plane maps are $N_s$ and $N_d$, respectively.

If the noise is Gaussian, we can write the likelihood of observing data $\mathbf{d}$ given a lensing operator $\mathbf{L}$ and source $\mathbf{s}$ as
\begin{equation}
\label{eq:chieq}
P(\mathbf{d} \; | \; \mathbf{L}, \mathbf{s}) \propto \exp \bigg(-E_\text{d}(\mathbf{d} \; | \; \mathbf{L}, \mathbf{s})\bigg),
\end{equation}
where
\begin{equation}
\label{eq:E_d}
E_\text{d}(\mathbf{d} \; | \; \mathbf{L}, \mathbf{s}) = \frac{1}{2}\chi^2(\mathbf{s}) = \frac{1}{2}(\mathbf{Ls-d})^\top \mathbf{C}_\text{d}^{-1}(\mathbf{Ls-d}),
\end{equation}
and $\mathbf{C}_\text{d}$ is the symmetric noise covariance matrix, which contains pixel-to-pixel noise correlations.
For the case of uniform, pixel-independent noise, $\mathbf{C}_\text{d}$ is diagonal with entries equal to $\sigma^2$, where $\sigma$ is the standard deviation of the noise.

\citet{suyureg} define the most likely solution, $\mathbf{s}_{\text{ml}}$, as the source model that maximises the likelihood and thus minimises $E_\text{d}$.
Setting $\nabla E_\text{d}(\mathbf{s}) = 0$, we find that $\mathbf{s}_\text{ml}$ satisfies
\begin{equation}
\label{eq:likelihoodnabla}
\mathbf{F}\mathbf{s} = \mathbf{f},
\end{equation}
where $\mathbf{F} = \mathbf{L}^\top \mathbf{C}_\text{d}^{-1} \mathbf{L}$ and $\mathbf{f} = \mathbf{L}^\top \mathbf{C}_\text{d}^{-1} \mathbf{d}$.
Because $\mathbf{F}$ is square and invertible by construction,\footnote{$\mathbf{F}^{-1}$ will fail to exist \CRKadd{if there are source pixels that cannot be constrained by the image pixels (i.e., there are too many source pixels overall, or source pixels that do not map to regions of the image plane with useful data), or if there are image pixels that lack corresponding source pixels. Those situations can generally be avoided with reasonable choices of grids.} It is conceivable that certain grid configurations could \CRKadd{also create problems for $\mathbf{F}^{-1}$}, but those should be rare.} $\mathbf{s}_\text{ml}$ is given by
\begin{equation}
\mathbf{s}_{\text{ml}}=\mathbf{F}^{-1}\mathbf{f}.
\label{eq:sml}
\end{equation}
If the left-inverse, $\mathbf{L}_\text{left}^{-1}$, of $\mathbf{L}$ exists,\footnote
{
 $\mathbf{L}_\text{left}^{-1}$ will exist and be unique if $\mathbf{L}$ is square and non-singular.
If $\mathbf{L}$ is rectangular, $\mathbf{L}_\text{left}^{-1}$ will exist if there are more image pixels than source pixels and $\mathbf{L}$ has full column rank.
These conditions may not be satisfied if two or more image pixels map to the same point (within machine precision), or if other similar coincidences occur.
}
 then Eq.~\ref{eq:sml} reduces to what one might na\"{\i}vely expect:
\begin{equation}
\mathbf{s}_{\text{ml}}=\mathbf{L}_\text{left}^{-1}\mathbf{d}.
\end{equation}

%%%%%%%%%%%%%%%%%%%%%%%%%%%%%%%%%%%%%%%%%%%%%%%%%%%%%%%%%%%%
\subsection{Most probable solution}
\label{sec:mps}

Unfortunately, $\mathbf{s}_{\text{ml}}$ will fit the noise in the data in addition to the lensed images.
There are several different ways to avoid such overfitting.
Maximum entropy methods \citep[MEMs;][]{citewallingtonMEM} favor sources whose pixel values follow broad distributions expected from information theory, as opposed to sources with some pixels that are very different from the rest.
MEMs also prohibit negative surface brightness values.
They do not constrain surface brightness variations between adjacent pixels, however, and can lead to large fluctuations over small scales.
To favor sources that are smooth, we might introduce a function that penalises large values of the first or second derivative \citep[the particular choice depends on the data and underlying source; see][]{citebrewer2006,suyureg}.
If the penalty function is quadratic in the source surface brightness, the source that maximises the likelihood while minimising the penalty is still given by a linear equation.

Suppose, for example, that we want to introduce a function $E_\text{s}(\mathbf{s})$ that penalises large surface brightness gradients.
We can define a derivative operator $\mathbf{H}$ that acts on a source vector $\mathbf{s}$ to produce a vector $\mathbf{Hs}$ containing the gradient of the surface brightness at each pixel.
Then we put
\begin{equation}
E_\text{s}(\mathbf{s}) = \frac{1}{2}(\mathbf{H}\mathbf{s})^\top\mathbf{H}\mathbf{s} = \frac{1}{2}\mathbf{s}^\top(\mathbf{H}^\top\mathbf{H})\mathbf{s} = \frac{1}{2}\mathbf{s}^\top\mathbf{R}\mathbf{s},
\end{equation}
where $\mathbf{R}\equiv\mathbf{H}^\top\mathbf{H}$.
In other words, when sandwiched between two source vectors, $\mathbf{R}$ returns the square of the gradient summed over source pixels.
A similar construction can return the sum of the squares of the curvature (see \S\ref{sec:reg}).

It is important to strike a balance between fitting the data and regularising the source (especially since any given regularisation scheme may not accurately represent the true source surface brightness).
This can be done by writing the full posterior probability distribution for the source model as
\begin{equation}
\label{eq:ppd4s}
P(\mathbf{s} \; | \; \mathbf{L}, \mathbf{R}, \mathbf{d}, \lambda) \propto \exp\bigg(-M(\mathbf{s})\bigg),
\end{equation}
where
\begin{equation}
M(\mathbf{s}) \equiv E_\text{d}(\mathbf{s}) + \lambda E_\text{s}(\mathbf{s}) ,
\label{eq:penalty}
\end{equation}
and $\lambda$ is a dimensionless parameter that determines which term in Eq.~\ref{eq:penalty} dominates.
When the ``regularisation strength'' $\lambda$ is small, the \CRKadd{Bayesian framework} will primarily fit the data; while when $\lambda$ is large, the \CRKadd{framework} will enforce strong priors on the source.

\citet{suyureg} define the most probable solution $\mathbf{s}_{\text{mp}}$ as the source model that maximises the posterior and thus minimises $M(\mathbf{s})$.
To find this model, we Taylor expand $E_\text{d}$ to second order about its minimum,
\begin{equation}
E_\text{d}(\mathbf{s}) = E_\text{d}(\mathbf{s}_{\text{ml}}) + \frac{1}{2}(\mathbf{s}-\mathbf{s}_{\text{ml}})^\top\mathbf{F}(\mathbf{s}-\mathbf{s}_{\text{ml}}).
\end{equation}
The matrix $\mathbf{F}$ that appears here is the Hessian\footnote{
The Hessian of a function is a matrix that contains the second order partial derivatives of the function.
In this case, the derivatives are taken with respect to the source vector.
For example, the $(i,j)$ entry of $\mathbf{F}$ would hold the second order derivative of $E_\text{d}$ with respect to the $i^{th}$ and $j^{th}$ source pixels.
}
of $E_\text{d}$, but from Eq.~\ref{eq:E_d} this is the same as $\mathbf{F}$ defined in Eq.~\ref{eq:likelihoodnabla}.
Setting $\nabla M(\mathbf{s}) = 0$, we find that $\mathbf{s}_{\text{mp}}$ satisfies
\begin{equation}
\label{eq:posteriornabla}
\mathbf{A}\mathbf{s} = \mathbf{F}\mathbf{s}_{\text{ml}},
\end{equation}
where $\mathbf{A} = \mathbf{F} + \lambda \mathbf{R}$ is the Hessian of $\mathbf{M}$ from Eq.~\ref{eq:penalty}.
Because $\mathbf{A}$ is square and invertible by construction, $\mathbf{s}_{\text{mp}}$ is given by
\begin{equation}
\label{eq:smp}
\mathbf{s}_{\text{mp}} = \mathbf{A}^{-1}\mathbf{F}\mathbf{s}_{\text{ml}} =\mathbf{A}^{-1}\mathbf{f}.
\end{equation}

It remains to determine the regularisation strength $\lambda$ seen in Eq.~\ref{eq:penalty}.
In the Bayesian framework, the optimal value of $\lambda$ is found by maximising
(see \citealt{suyureg} for a full discussion)
\begin{equation}
P(\lambda \; | \; \mathbf{d}, \mathbf{L}, \mathbf{R}) \propto P(\mathbf{d} \; | \; \mathbf{L}, \lambda, \mathbf{R})P(\lambda).
\end{equation}
We assume a uniform logarithmic prior, $P(\lambda) \propto \lambda^{-1}$, because we do not know the scale of $\lambda$ a priori.
The optimal regularisation strength, $\hat{\lambda}$, can then be found numerically.

Formally, $\mathbf{s}_\text{mp}$ is a biased estimator of the true source surface brightness $\mathbf{s}_\text{true}$. \citet{suyureg} show that averaging over many realisations yields
\begin{equation}
\langle \mathbf{s}_\text{mp}\rangle = \mathbf{A}^{-1}\mathbf{F}\mathbf{s}_\text{true} ,
\label{eq:biassmp}
\end{equation}
which differs from $\mathbf{s}_\text{true}$ to the extent that $\mathbf{A}^{-1} = (\mathbf{F} + \lambda \mathbf{R})^{-1}$ differs from $\mathbf{F}^{-1}$.
The simulations presented in \S\ref{sec:mockdatab} allow us to quantify the extent to which the bias translates into errors on recovered lens model parameters.

%%%%%%%%%%%%%%%%%%%%%%%%%%%%%%%%%%%%%%%%%%%%%%%%%%%%%%%%%%%%
\subsection{Model ranking}

Once we solve for the source at a fixed lens model, we must rank different models by evaluating the posterior probability
\begin{equation}
\label{eq:evieq}
P(\mathbf{L},\mathbf{R} \; | \; \mathbf{d}) \propto P(\mathbf{d} \; | \; \mathbf{L},\mathbf{R}) \times \text{priors on } \mathbf{L} \text{ and } \mathbf{R}.
\end{equation}
If the priors on the lens models and regularisation scheme are flat, then we can just evaluate the Bayesian evidence\footnote{
Strictly speaking, this is not the full evidence because the lens model parameters are not marginalised, but the terminology is standard.
}
$P(\mathbf{d} \, | \, \mathbf{L},\mathbf{R}) = \int P(\mathbf{d} \, | \, \mathbf{L}, \lambda, \mathbf{R}) \, P(\lambda) \, \mathrm{d}\lambda$.
\citet{suyureg} \CRKadd{suggest} that the distribution for $\lambda$ \CRKadd{can be expected to have a sharp peak}, so instead of computing the full integral we can just evaluate the integrand at its peak.

Examining Eqs.~\ref{eq:chieq} and \ref{eq:evieq}, we can infer that
\begin{equation}
\label{eq:lasteqbayesian}
-2\ln \mathcal E = \chi^2 + V,
\end{equation}
where $\mathcal E$ is shorthand for the evidence and $V$ is a constant that depends on the available prior volume of the parameter space.
Thus, we will use $\chi^2$ and $-2\ln \mathcal E$ interchangeably.
For a more detailed discussion on the connection between evidence and $\chi^2$, see \citet{cite:chieviconnect}.

%%%%%%%%%%%%%%%%%%%%%%%%%%%%%%%%%%%%%%%%%%%%%%%%%%%%%%%%%%%%
\section{Test data}
\label{sec:test_data}
%%%%%%%%%%%%%%%%%%%%%%%%%%%%%%%%%%%%%%%%%%%%%%%%%%%%%%%%%%%%

\begin{figure}
\centering
\includegraphics[width=0.45\textwidth]{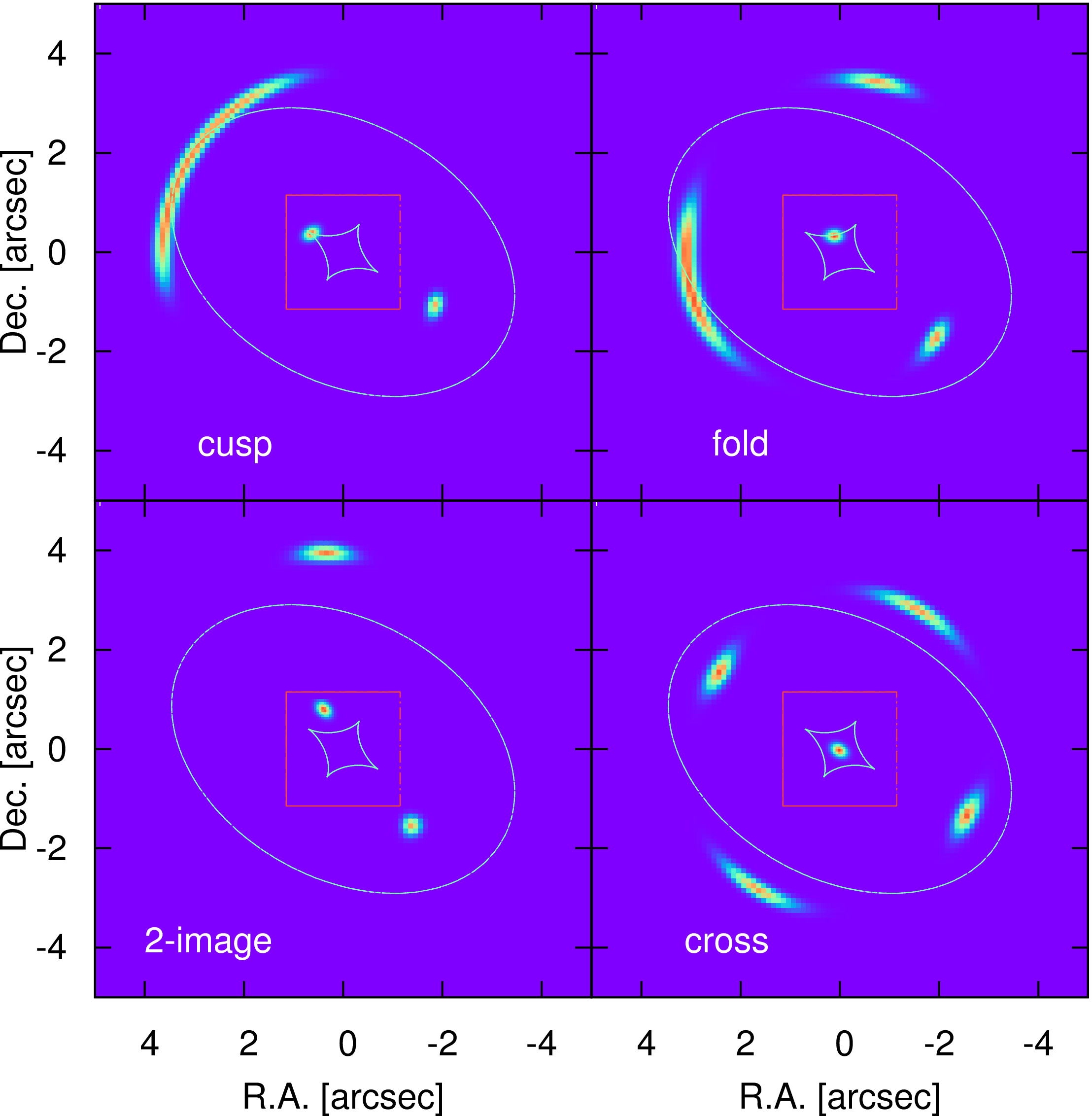}
\caption{\small
Test data in four canonical configurations.
Clockwise from top left: cusp, fold, cross, and 2-image lens configurations.
For each panel, the diamond-shaped curve (the caustic) and object (source galaxy) inside the red square are in the source plane, while the elliptical curve (the critical curve) and the other features (the arcs) outside the red square are in the image plane.
The lensing galaxy used to create the data is the same in all cases: a SIE with Einstein radius of $3''$, ellipticity of $0.3$, position angle of the semi-major axis $60^\circ$ east of north.
The colour scale is linear and identical in all panels.
The source galaxies used to create the data share the same size and luminosity profile; only the positions and orientations differ.}
\label{fig:mockdata}
\end{figure}

\begin{figure}
\centering
\includegraphics[width=0.45\textwidth]{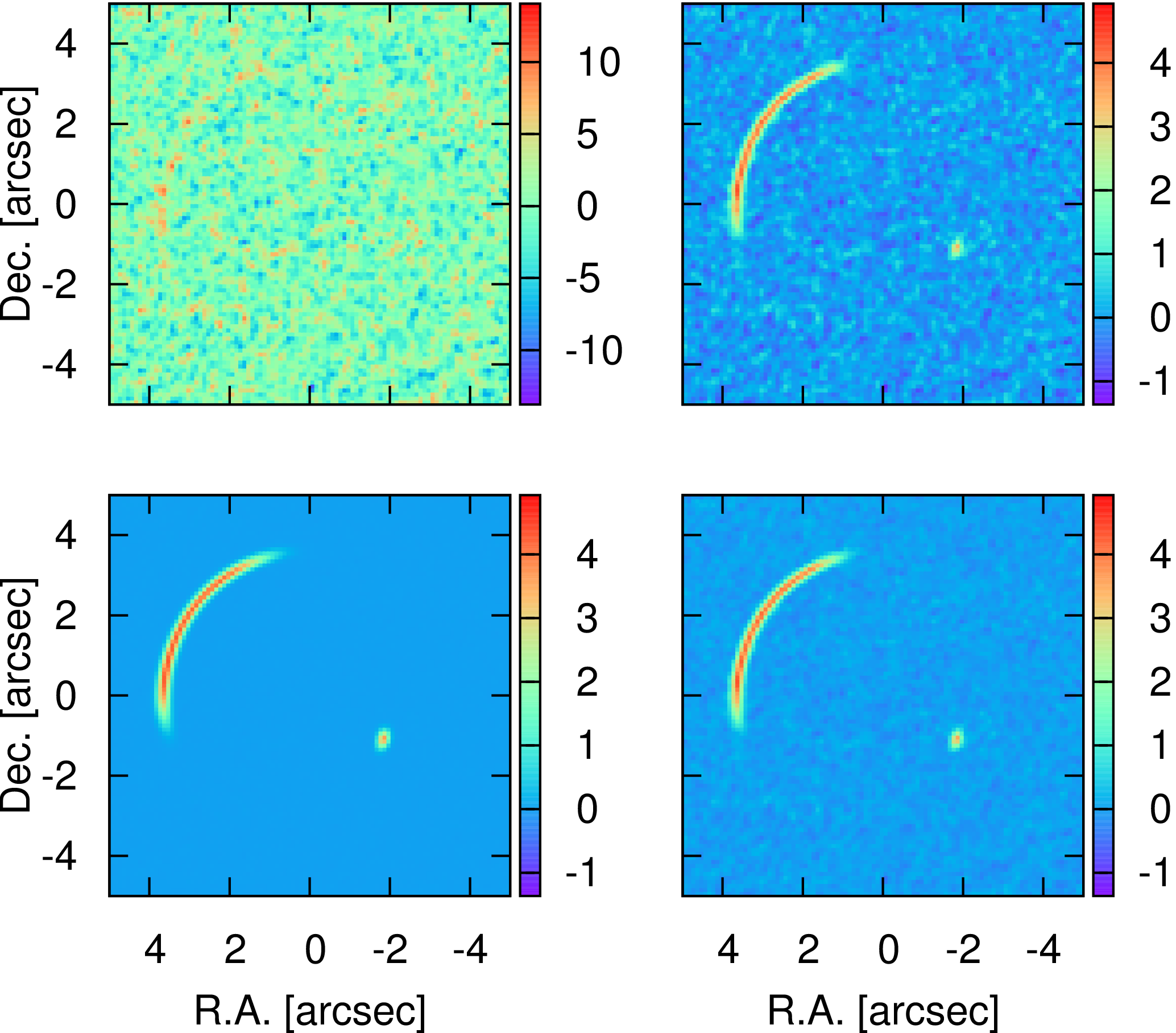}
\caption{\small
Test data for the cusp configuration shown with varying noise levels.
Peak S/N clockwise from top left: 1, 10, 25, 500.
The pixel scale is 0.1 arcsec/pixel.
The colour scale is linear and consistent except for the $\text{S/N}=1$ case.}
\label{fig:mockdatanoise}
\end{figure}

\begin{figure}
\centering
\includegraphics[width=0.45\textwidth]{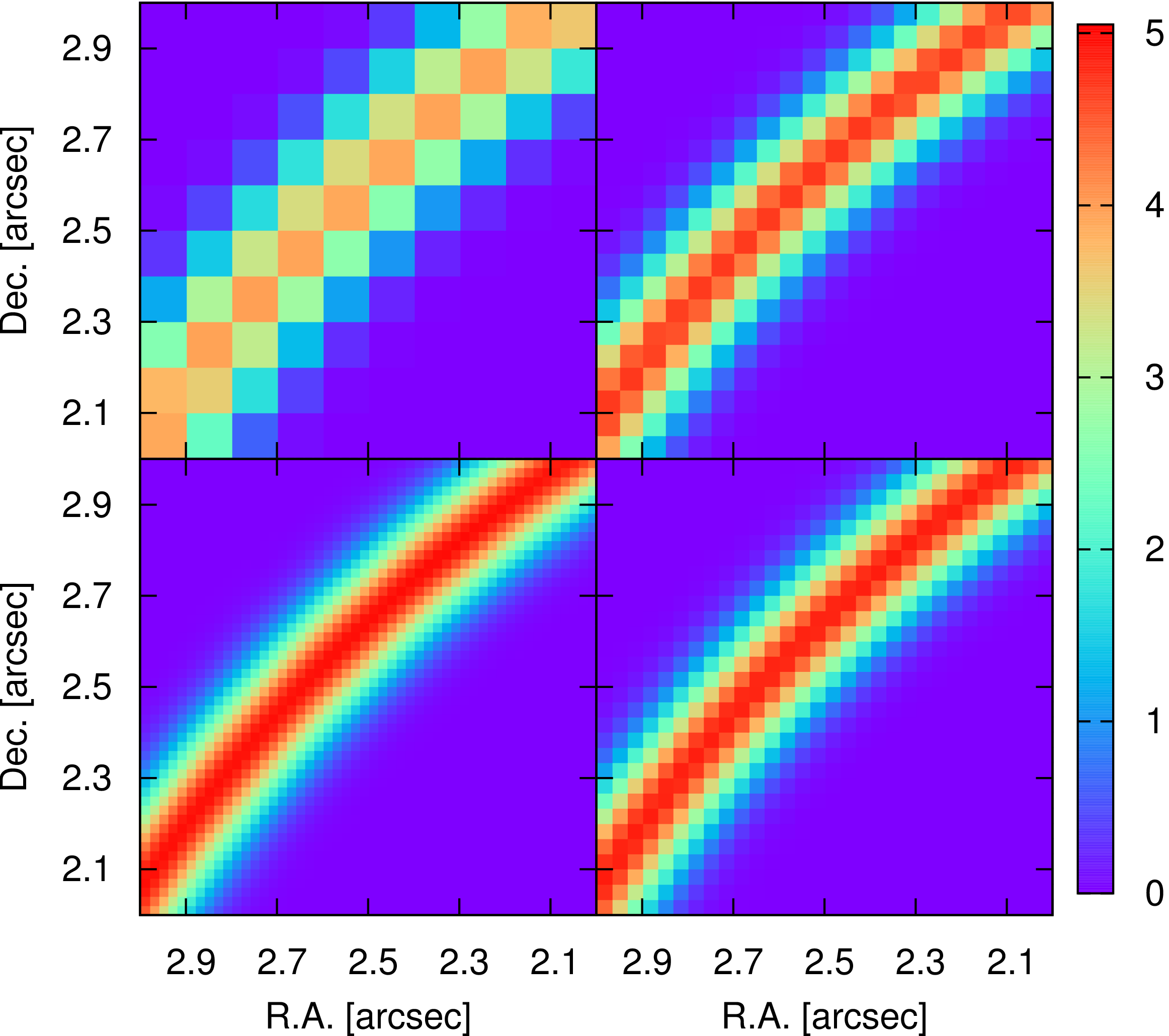}
\caption{\small
Test data for the cusp configuration shown with varying pixel scales.
Only a subsection of the long arc is shown so that the differences in resolution are visible.
Clockwise from top left: 0.1, 0.05, 0.03, and 0.02 arcsec/pixel.}
\label{fig:mockdatares}
\end{figure}

To explore possible uncertainties and biases in PBSR algorithms, we construct test data using a simple but realistic lens and source.
The lens is a singular isothermal ellipsoid (SIE), which is a popular choice for modeling elliptical galaxies.
Although the dark and luminous mass profiles are not simple power laws individually, the total density profile appears to be close to isothermal \citep{isothermalconspiracy,koopmansisothermal,2010ARAA}.
The SIE is placed at the origin and fixed with an Einstein radius of $3''$, ellipticity of $0.3$, and position angle of $60^\circ$ east of north.
The source luminosity profile is an elliptical Gaussian with a half-light radius of $0.125''$ and peak surface brightness of 5 (in arbitrary units).
The position and orientation of the source are varied to create four canonical lens configurations that let us assess whether uncertainties in PBSR algorithms are sensitive to the image morphology.
Fig.~\ref{fig:mockdata} shows a 2-image configuration along with three configurations that nominally have four images: a source near the center of the caustic produces a ``cross'' configuration with four distinct images; a source just inside the caustic curve produces two short arcs and a long arc from two merging images (a ``fold'' configuration); and a source inside a cusp in the caustic produces one long arc from three merging images along with an isolated image on the other side (a ``cusp'' configuration).
In the following sections we vary the amount of noise in the mock data (Fig.~\ref{fig:mockdatanoise}) and the resolution (i.e., the pixel scale; Fig.~\ref{fig:mockdatares}).

%%%%%%%%%%%%%%%%%%%%%%%%%%%%%%%%%%%%%%%%%%%%%%%%%%%%%%%%%%%%
\section{Issues intrinsic to the algorithm}
%%%%%%%%%%%%%%%%%%%%%%%%%%%%%%%%%%%%%%%%%%%%%%%%%%%%%%%%%%%%

Some of the practical challenges in PBSR are inherent to the algorithm itself.
We have already mentioned the need for regularisation.
Dealing with gridded data makes some degree of interpolation unavoidable.
Also, different parts of the image plane probe different spatial scales in the source plane, depending on the lensing magnification.
Using an adaptive source plane grid helps take full advantage of the information contained in a lensed image, but leads to challenges with interpolating and calculating derivatives on an irregular grid.
In this section we examine how these issues affect the source reconstruction and lens model ranking.

%%%%%%%%%%%%%%%%%%%%%%%%%%%%%%%%%%%%%%%%%%%%%%%%%%%%%%%%%%%%
\subsection{Gridding}
\label{gridsection}

In PBSR, the image and source grids do not have to be the same.
Image pixels have definite dimensions set by the instrument and data processing.
But source ``pixels'' are more general; they refer loosely to positions (and small regions around them) where one chooses to reconstruct the surface brightness of the source.
The shape and density of source pixels are arbitrary, and they can vary across the source plane.
The source pixel density directly limits the resolution of the reconstruction.

If source pixels outnumber image pixels, the reconstruction problem will be underconstrained.
The regularisation strength will be driven to high values, effectively decreasing the number of independent source pixels.\footnote{
Strong regularisation introduces correlations between nearby pixels, smoothing the surface brightness and decreasing the effective resolution in the source plane.
}
In each of the gridding schemes discussed below, the grid is constructed so the number of source pixels is approximately half the number of image pixels.

The size and shape of the grid can be limited to specific regions on the sky.
Using all image pixels may be computationally expensive, and it can make the regularisation less effective (because most of the source pixels would just contain noise).
Therefore it may be useful to construct masks around regions that contain lensed images.
\citet{citewaythmask} comment on the importance of careful pixel masking, because pixels that do not contain flux can be as important as those that do.
If a model fits the observed surface brightness but also puts flux where no light is observed, the model should be penalised but overly aggressive masking might cause the  faulty pixels to be ignored.
As a precaution, \textit{pixsrc} can find and include all pixels that are ``sisters'' to the pixels in the masked region(s).\footnote{Heuristically, image pixels are sisters if they come from the same source pixel.}
Doing so requires some care because the number of image pixels that get used can vary with the lens model.

\begin{figure}
  \centering
  \begin{tabular}{ c }
    \includegraphics[width=0.45\textwidth]{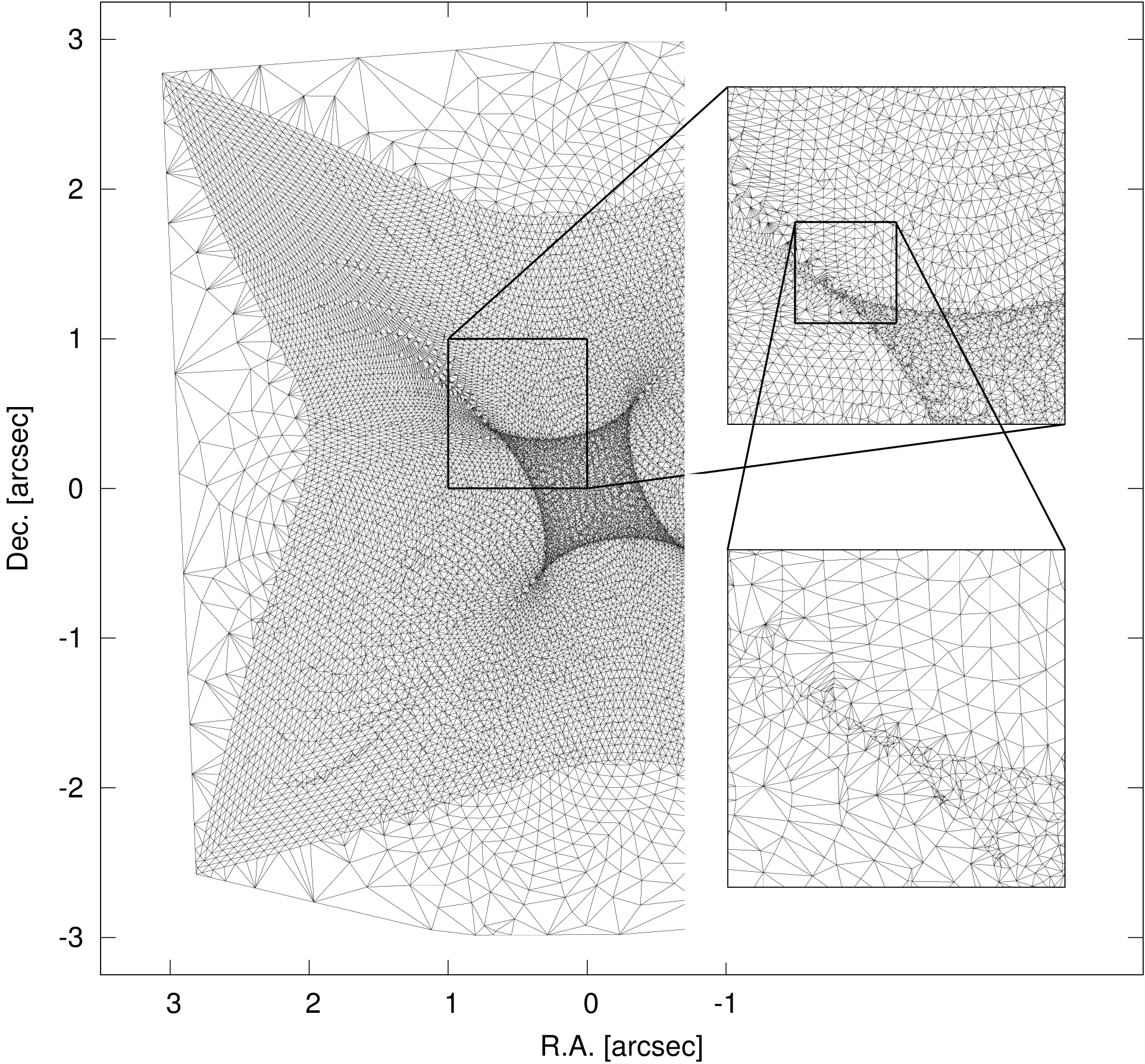} \\
    \includegraphics[width=0.45\textwidth]{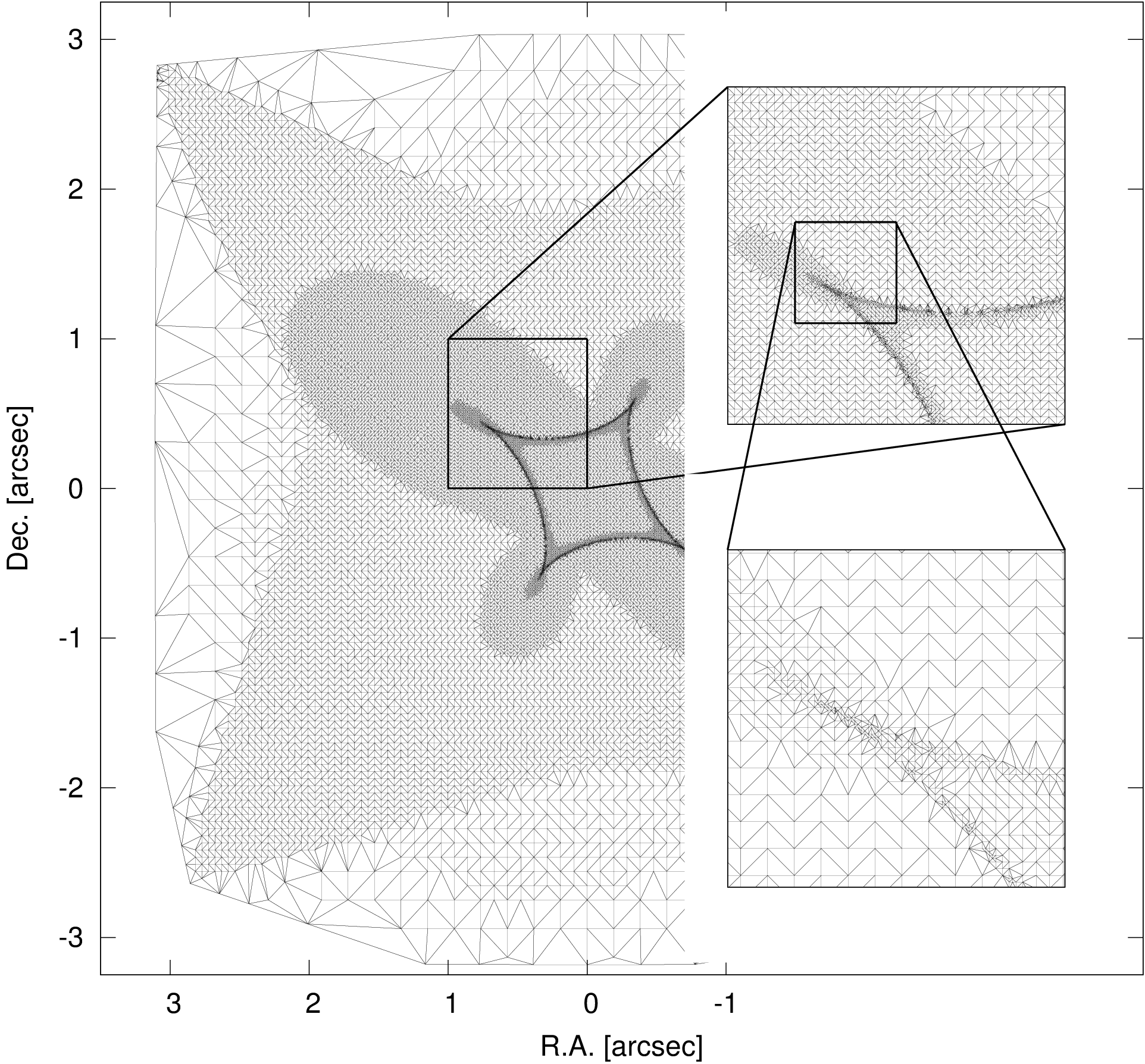}
  \end{tabular}
  \caption{\small
Triangulation of a fully adaptive grid (top) and an adaptive Cartesian grid (bottom).
The lens model used is specified in \S\ref{gridsection}: a SIE located at the origin with Einstein radius $3''$, ellipticity 0.3, and position angle $60^\circ$ east of north.}
\label{fig:grid}
\end{figure}

We describe three different schemes for gridding the source plane: one Cartesian and two adaptive.
Fig.~\ref{fig:grid} shows examples of the two adaptive grids.
We compare the performance of the two adaptive gridding schemes in \S\ref{sec:obstacles}.

\subsubsection{Cartesian grid}
\label{sec:grid-Cartesian}

We begin with a simple Cartesian grid.
The pixel density and resolution in the source plane are uniform.
The grid dimensions and pixel scale can be set manually or chosen to achieve $N_\text{s} \approx N_\text{d}/2$, as this seems to adequately reconstruct the source without being underconstrained.
Benefits of the Cartesian grid lie in its simplicity: the grid, lensing operator, and regularisation operator are easily and quickly constructed.
However, the uniform resolution means that small scales cannot be probed without incurring a large number of source pixels and a correspondingly large regularisation strength.

\subsubsection{Fully adaptive grid}
\label{sec:grid-FA}

\begin{figure}
\centering
\includegraphics[width=0.45\textwidth]{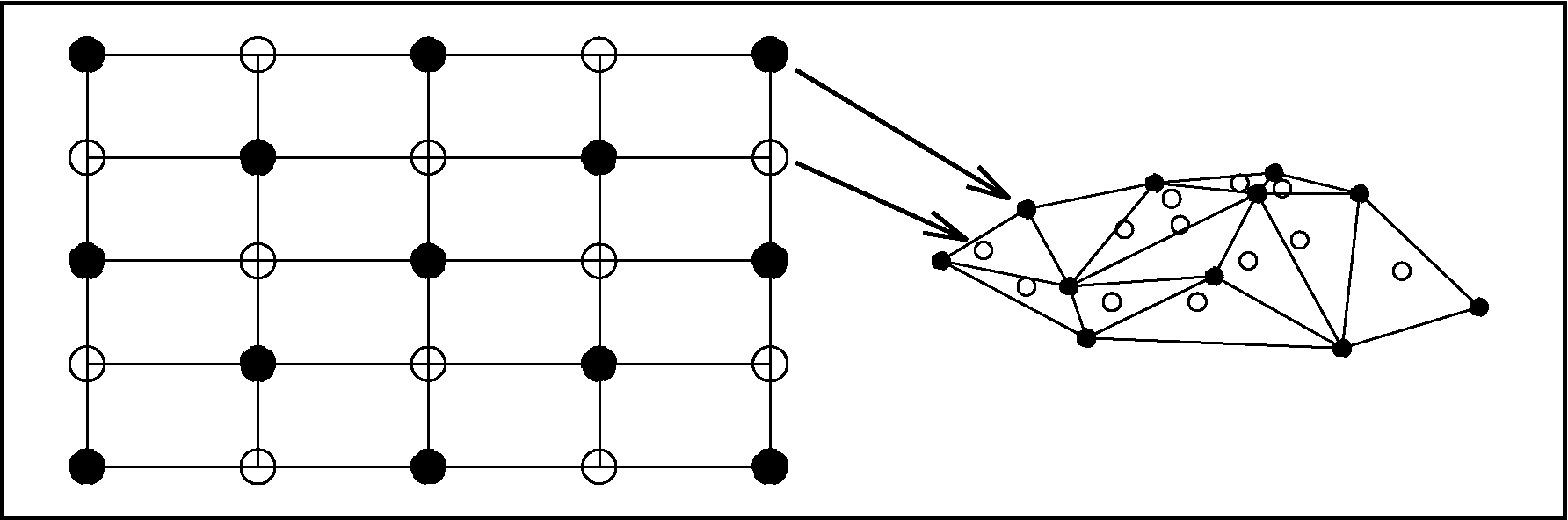}
\caption{\small
A demonstration of the fully adaptive grid construction.
The left and right panels show the image and source grids, respectively.
Filled circles in the image plane are mapped to the source plane, and a Delaunay triangulation \citep{citetriangle} is used to construct the source grid.
Open circles in the image plane are then mapped to the source plane and set to values interpolated from the surrounding source pixels.
Each filled circle has a row in the unblurred lensing operator with a single entry of 1, while each open circle has a row with three non-negative entries that sum to 1.
This figure is inspired by Fig. 1 in \protect\citet{vegettigrid}.}
\label{fig:vegettigridmaking}
\end{figure}

\citet{vegettigrid} introduced a gridding scheme in which some of the image plane pixels are mapped to the source plane and used to construct the source grid (see Fig.~\ref{fig:vegettigridmaking}).
By default we choose to use every other pixel to construct the grid, which helps to ensure that $N_\text{s} \approx N_\text{d}/2$.
The advantage of this ``fully adaptive'' grid is that the density of pixels in the source plane is set directly by the lens mapping, so it automatically achieves the natural resolution of lensing.
The challenge is that computing the derivatives needed for regularisation can be difficult on an irregular grid (see \S\ref{sec:reg} and Fig.~\ref{fig:reg0}).

\subsubsection{Adaptive Cartesian grid}
\label{sec:adacargri}

The adaptive Cartesian grid builds from the Cartesian grid.
An initial two-dimensional grid is refined, adding or removing pixels, so the pixel density varies according to some criterion.
Such adaptive mesh refinement algorithms have been used in many fields of research, including star formation modeling, radiative transfer codes, and magnetohydrodynamic simulations.
For PBSR, we implement an adaptive Cartesian grid similar to that used by \citet{dyegrid}, which is designed to place more source pixels in regions of higher magnification.
We first give a heuristic description of the gridding scheme, and then provide more details.

An initial, zeroth level grid is constructed as a box just large enough to contain all of the ray-traced image pixels, with five grid points (at the corners and center).
A zeroth level magnification, $\mu_0$, is ascribed to this grid.
Then each quadrant is examined, and if the magnification in this quadrant, $\mu_1$, is larger than four times the magnification of the parent grid ($\mu_1 \geq 4 \mu_0$), the quadrant is split into a (first level) subgrid, itself consisting of four quadrants.
The factor of four here is necessary because as we add a subgrid, we split a quadrant into four more quadrants, increasing the spatial resolution by a factor of four (in area).
Then, for each of these first level quadrants, we add a second level of subgridding if $\mu_2 \geq 4 \mu_1 = 16 \mu_0$.
This process is repeated for every quadrant and subquadrant.

In practice, it would be computationally expensive to examine every quadrant and subquadrant, and it would be undesirable to do so since many source pixels would be unused in the lensing operator.
Instead, every image pixel is ray-traced back to the source plane, the local magnification at that location is computed, the appropriate level of subgridding is determined based on the ratio of the local magnification to the zeroth level magnification, and only the minimum number of source pixels (three or fewer) needed are created.

It still remains to determine $\mu_0$.
Because the size of the zeroth level grid is arbitrary, $\mu_0$ is also arbitrary.
This freedom is what \citet{dyegrid} encapsulate in their ``splitting factor.''
We note that \citet{dyegrid} allow their splitting factor to vary in the source reconstruction.
We have not explored this additional freedom.
Instead, we fix $\mu_0$ so that $N_\text{s} \approx N_\text{d}/2$.

The appeal of the adaptive Cartesian grid lies in its use of the magnification as a physical motivation for adaptive gridding.
As we will see in \S\ref{sec:obstacles} and Fig.~\ref{fig:1devicut}, the noise in the $\chi^2$ surface is larger using the adaptive Cartesian grid.
The higher noise is thought to be due to the discrete change in magnification required to trigger the subgridding.
However, as discussed in \S\ref{sec:reg} and Fig.~\ref{fig:reg0}, derivatives seem to be computed more accurately.

%%%%%%%%%%%%%%%%%%%%%%%%%%%%%%%%%%%%%%%%%%%%%%%%%%%%%%%%%%%%
\subsection{Interpolation}
\label{sec:sinparvar}

The surface brightness of an image pixel is calculated by ray-tracing the pixel to the source plane and linearly interpolating over up to three adjacent source pixels.
Such interpolation amounts to treating the source as a collection of small planes, which may or may not provide an accurate approximation to the true surface brightness distribution (depending on the pixel scale).
Errors from the interpolation can be important if they are large compared with the random noise in the data.

\begin{figure}
\centering
\includegraphics[width=0.45\textwidth]{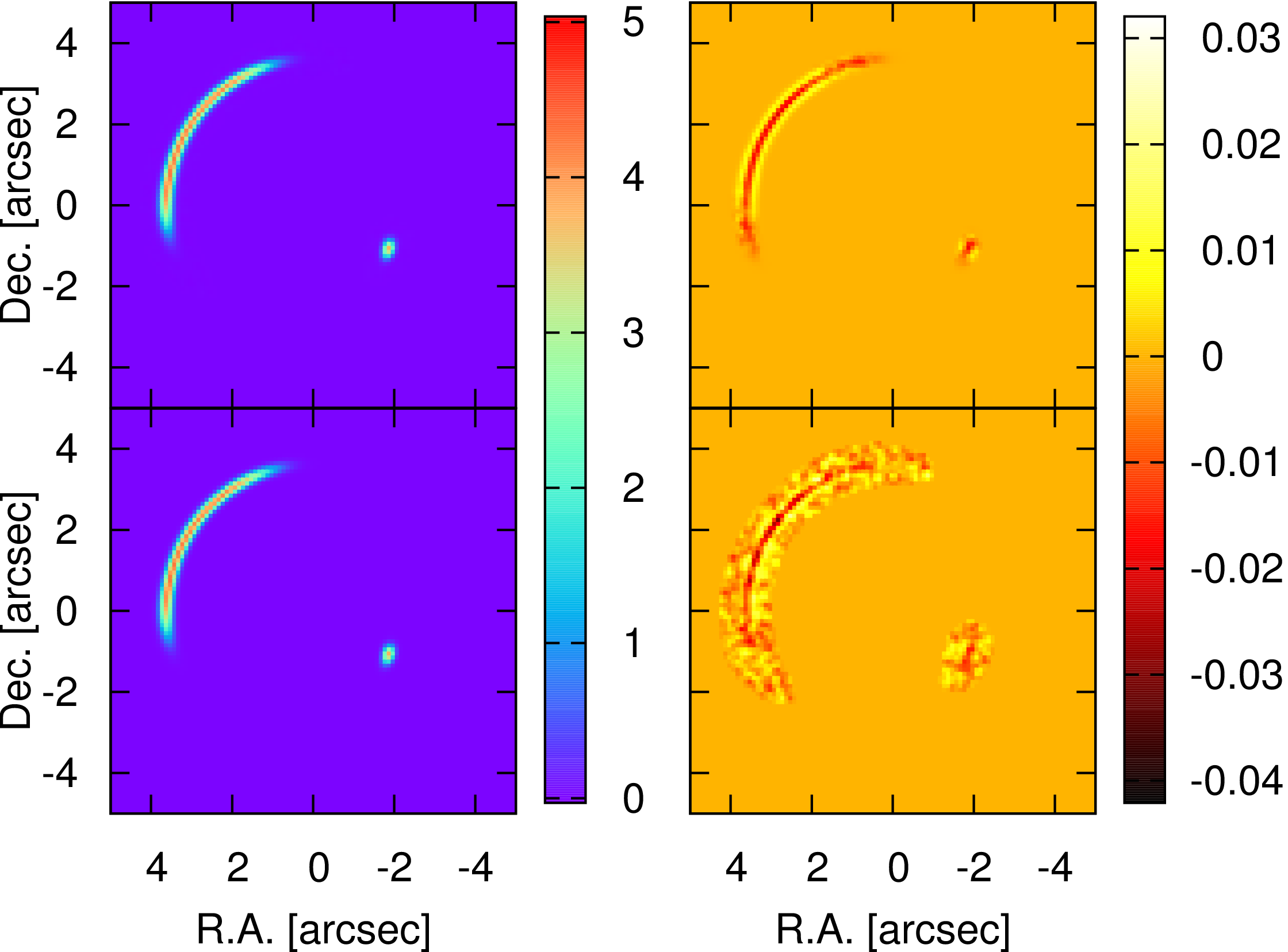}
\caption{\small
Visualisation of interpolation errors for a source in the cusp configuration with a peak S/N of 500 and a resolution of 0.03 arcsec/pixel.
From top left, clockwise: data, interpolation errors, model residuals, model.
The lens and source models were fixed at their correct values, but there are significant residuals with the same structure as the interpolation errors.
Accounting for interpolation errors lowers the $\chi^2$ from 13,236 to 8275, corresponding to a change in reduced $\chi^2$ from 1.60 to 0.999.
}
\label{fig:interperr2}
\end{figure}

As an illustration, Fig.~\ref{fig:interperr2} shows data, model, and residuals for a high-quality image of a source in the cusp configuration.
The peak S/N is 500, and the image resolution is 0.03 arcsec/pixel.
The lens model was fixed at the correct model, and the fully adaptive grid was constructed as usual, but the source surface brightness was fixed at the known value for the cusp source (rather than being reconstructed).
By visual inspection, the data and model seem to agree well, but the residuals show clear structure.
Also, the $\chi^2$ value is 13,236, which corresponds to a reduced $\chi^2$ of 1.60.
This is troubling since the lens and source models were fixed at the true values.
The residuals, and hence the large $\chi^2$ value, arise from interpolation errors.
To see this, Fig.~\ref{fig:interperr2} shows the difference between an image constructed directly from the analytic source and an image constructed from the interpolated version.
The structure of the interpolation errors clearly explains the structure of the model residuals.

We need to find a way to account for these errors.
Strictly speaking, we would have to know the true surface brightness of the source in order to determine interpolation errors in the first place.
As an approximation, we fit an analytic model (comprising one or more S\'{e}rsic profiles) to the pixelated source.\footnote{If the fit to the pixelated source is poor, we do not account for interpolation errors.}
We use this analytic model to compute a map of interpolation errors, as shown in the top right panel of Fig.~\ref{fig:interperr2}.
(The error map can be blurred by the PSF as needed.)
We then modify the noise covariance matrix ($\mathbf{C}_\text{d}$ in Eq.~\ref{eq:E_d}) with the substitution
\begin{equation}
\mathbf{C}_\text{d} \rightarrow \mathbf{C}_\text{d} + \mathbf{C}_\text{interp}.
\end{equation}
We make $\mathbf{C}_\text{interp}$ a diagonal matrix containing the squares of the interpolation errors (which omits any correlations in errors among pixels but is a simple and effective approach). This modification lowers the $\chi^2$ value for the case shown in Fig.~\ref{fig:interperr2} to 8275, which corresponds to a reduced $\chi^2$ of 0.999.

\begin{figure}
\centering
\includegraphics[width=0.45\textwidth]{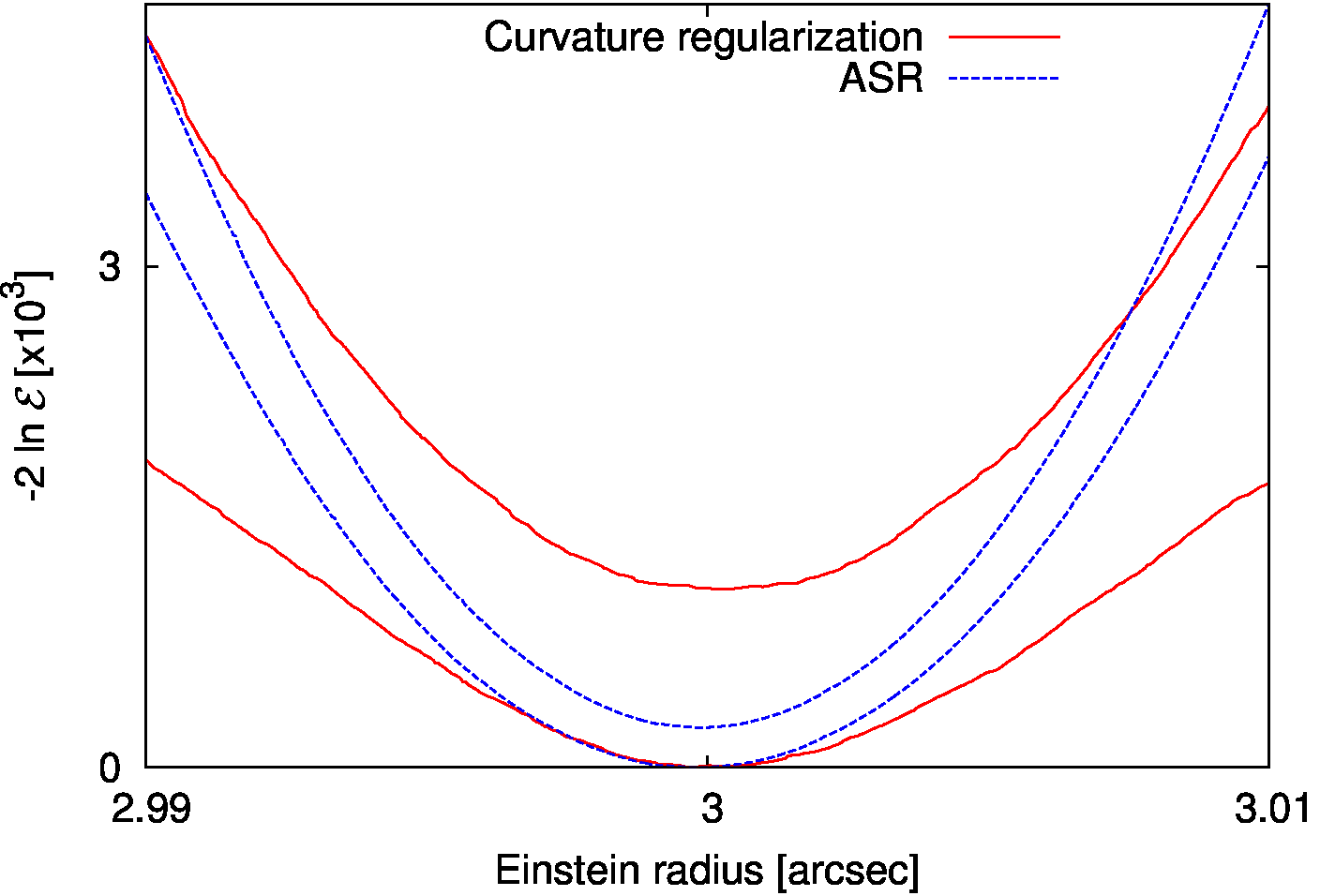}
\caption{\small
Effect of interpolation errors, when the errors are comparable to the noise level.
The test data have a source in the cusp configuration with a peak S/N of 100 and a resolution of 0.05 arcsec/pixel.
Red, solid lines and blue, dashed lines correspond to curvature regularisation and ASR, respectively (see \S\ref{sec:reg} for a discussion of regularisation schemes).
The upper lines do not account for interpolation errors, while the lower lines do.
Because we focus on differences in $\chi^2$, vertical offsets have been applied, but differences between same colour curves are meaningful.
Qualitatively, we see that accounting for interpolation errors broadens the $\chi^2$.
Quantitatively, the ranges of $\chi^2$ change by factors of 1.8 and 1.2 for curvature regularisation and ASR, respectively.
}
\label{fig:interperr}
\end{figure}

Accounting for interpolation errors in this way is a conservative practice, as the effect is to broaden the $\chi^2$ surface.
As an example, Fig.~\ref{fig:interperr} shows a one dimensional cut of the Bayesian evidence for a cusp configuration with a pixel scale of 0.05 arcsec/pixel and a peak S/N of 100.
All lens model parameters except the Einstein radius are fixed at their true values.
The curves show the Bayesian evidence as a function of $R_E$ for two forms of regularisation (discussed in \S\ref{sec:reg}), when we do or do not account for interpolation errors.
Although the location of the minimum does not appear to change, the $\chi^2$ curve becomes shallower when interpolation errors are addressed, reflecting a larger uncertainty in the Einstein radius.

\begin{figure}
\centering
\includegraphics[width=0.45\textwidth]{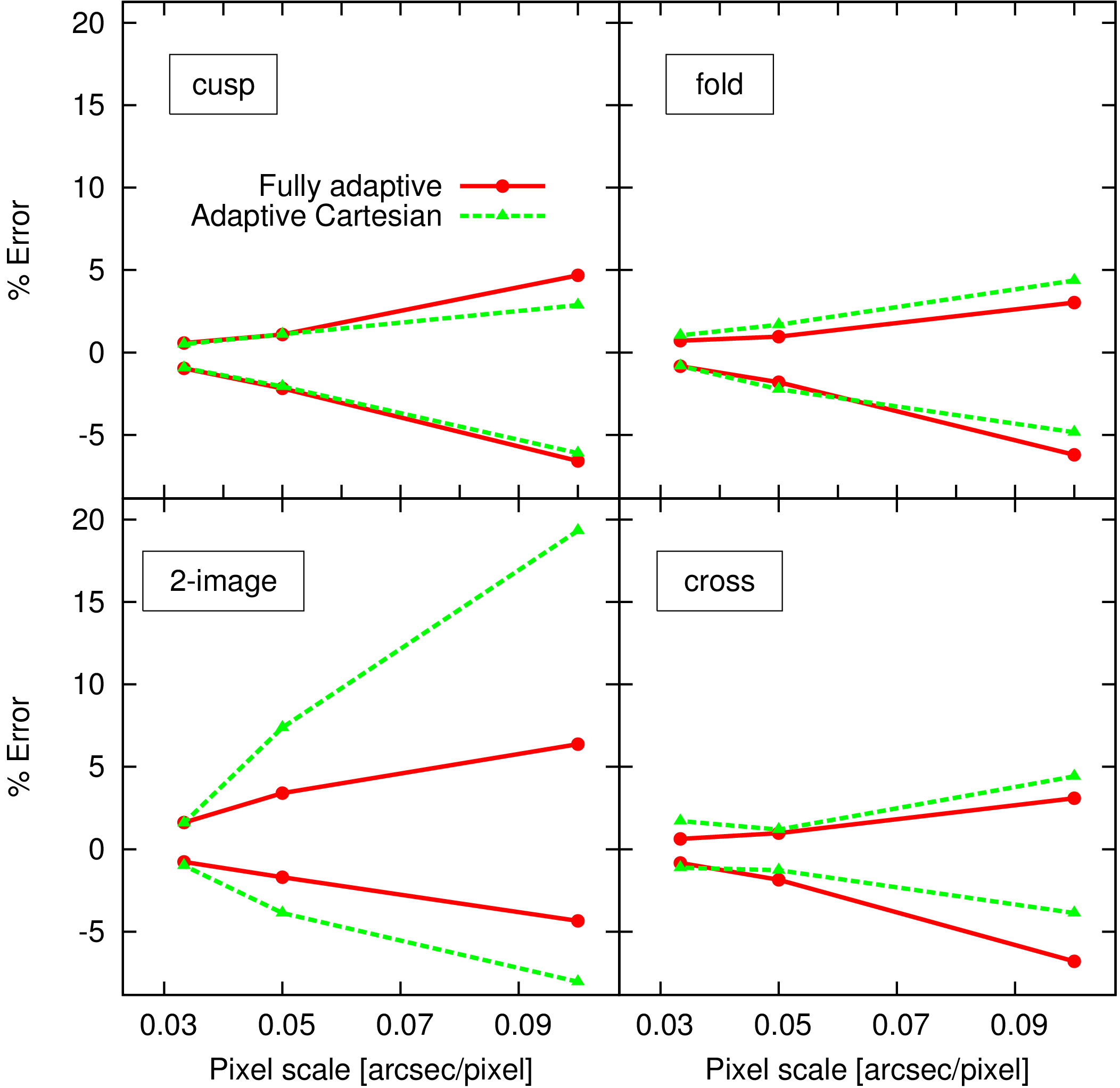}
\caption{\small
Minimum and maximum interpolation errors are shown as a function of the pixel scale for both gridding schemes.
(The pixel scale is quoted for the image plane because that is the quantity known from data, but bear in mind that interpolation occurs in the source plane.)
From top left, clockwise: cusp, fold, cross, and 2-image configurations.
The lens and source models were fixed at their correct values.
The percent error in any given image pixel is calculated by taking the ratio of the interpolation error in that pixel to the peak signal in the image.
The doubly-imaged configuration shows the largest errors, because the magnification is lower and hence the number of source pixels that cover the source is smaller.
}
\label{fig:interperr3}
\end{figure}

The scale of interpolation errors depends on the lens configuration and image resolution.
Fig.~\ref{fig:interperr3} shows the minimum and maximum interpolation errors as a function of the pixel scale for all four lens configurations, using both the fully adaptive and adaptive Cartesian grids.
The lens model and source brightness are again fixed at their true values.
As the image resolution improves, the interpolation errors decrease.
The doubly-imaged source is not as highly magnified as the other cases, so the effective resolution in the source plane is lower and the interpolation errors are larger (reaching about 20\% of the peak flux).
The quad configurations show interpolation errors up to about 7\%.

%%%%%%%%%%%%%%%%%%%%%%%%%%%%%%%%%%%%%%%%%%%%%%%%%%%%%%%%%%%%
\subsection{Regularisation}
\label{sec:reg}

In \S\ref{sec:mps} we discussed regularising the source by penalising large values of the first or second derivative of the surface brightness distribution. In this section we explore two methods for computing the required numerical derivatives: a finite difference method (FDM) and a divergence theorem method (DTM).
\CRKadd{Note that the formulae in this section are deliberately written so that each source pixel receives equal weight in the regularisation; the formulae would have to be modified to weight pixels by the area they subtend in order to obtain true derivatives of the source surface brightness distribution.
We use equal weighting to take advantage of the fact that lensing effectively gives higher resolution in regions that are more highly magnified (see \citealt{vegettigrid} for more discussion).
This choice makes the regularisation sensitive to the lens model through the density of source pixels, so in principle it might introduce model-dependent biases into the regularisation.
The simulations presented in \S\ref{sec:mockdatab} suggest that such biases are small in practice.}

\begin{figure}
\centering
\includegraphics[width=0.45\textwidth]{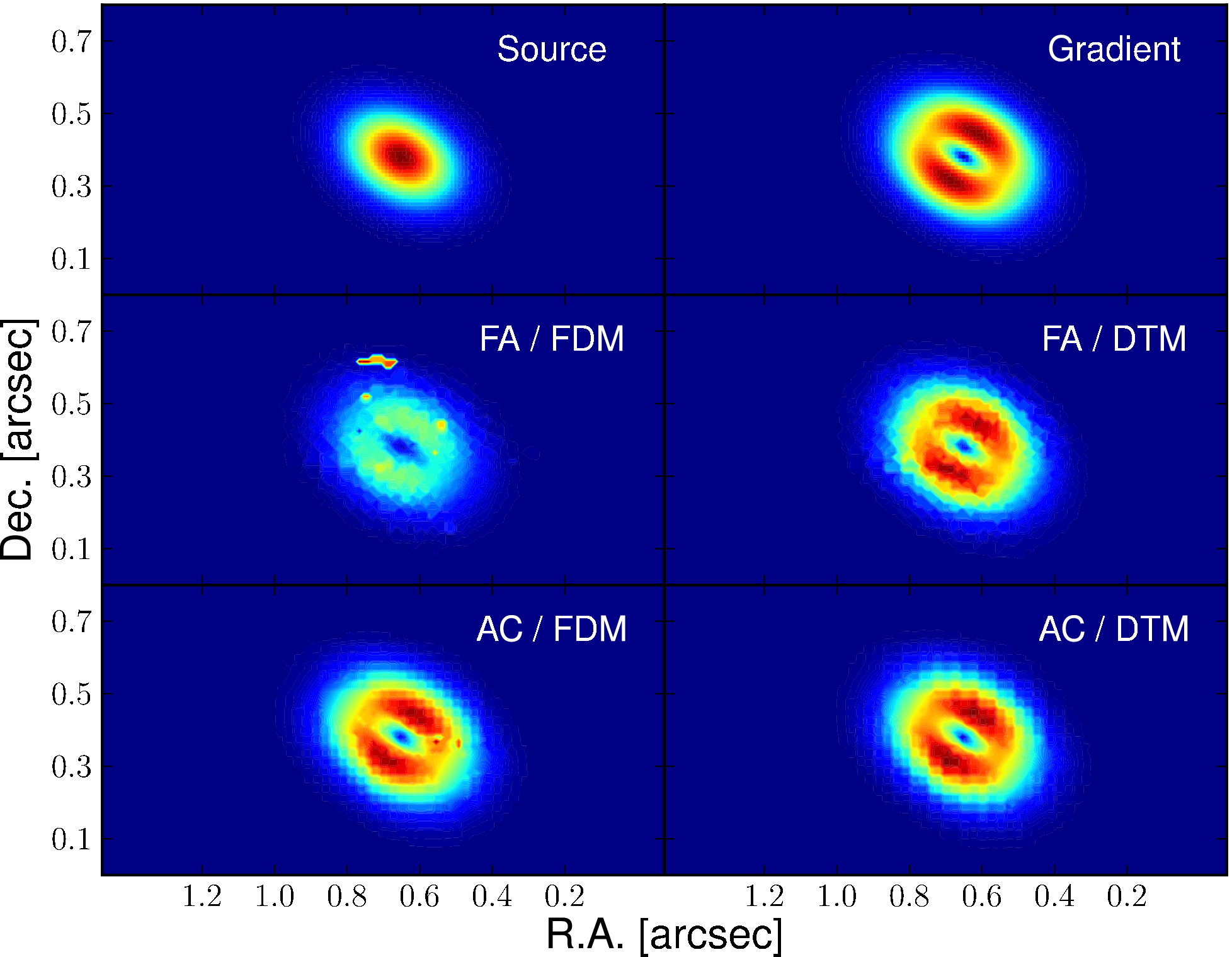}
\caption{\small
Comparison of FDM and DTM on both adaptive grids.
Top left: fixed source model (an elliptical Gaussian with ellipticity 0.3), placed in the cusp configuration.
Top right: exact magnitude of the gradient of source.
The remaining panels show the magnitude of the gradient computed with various grids and derivative schemes.
The middle row corresponds to the fully adaptive (FA) grid, while the bottom row corresponds to the adaptive Cartesian (AC) grid.
The left column corresponds to FDM, while the right column corresponds to DTM.
The colour scale is linear.
Only relative changes within a panel are important, because multiplicative constants can be absorbed into the regularisation strength.
The spurious peaks sometimes seen when using the FDM are likely due to  unfortuitous alignment of ``virtual pixels'' with the pixel at which the derivative is evaluated. For a more detailed discussion, see \S\ref{sec:fdm} and \S\ref{sec:dtm}.
}
\label{fig:reg0}
\end{figure}

Fig.~\ref{fig:reg0} \CRKadd{shows how the two methods perform} on both the fully adaptive and adaptive Cartesian grids.
The source is an elliptical Gaussian with ellipticity 0.3, and the magnitude of the gradient is shown.
It is important to note that only relative magnitudes are meaningful, because the regularisation strength can absorb multiplicative factors.
For the fully adaptive grid, DTM yields much better results.
For the adaptive Cartesian grid, the difference is less significant but there is still some improvement going from FDM to DTM.

We also introduce a regularisation scheme that penalises the source model for deviations from an analytic source profile and refer to this method as analytic source regularisation (ASR).

\subsubsection{Finite difference method}
\label{sec:fdm}

Using Taylor's theorem, we can calculate derivatives on a grid using the finite difference method (FDM).
For a simple Cartesian grid, the gradient at a particular pixel $m$ can be approximated by taking directional finite differences of the surface brightness along the grid axes at the pixel $m$. This can be written as
\begin{equation}
\label{firstdereq}
\begin{aligned}
\vec{\mathbf{g}}[m] &= \frac{1}{2} \sum_n \bigg(\mathbf{s}[n]-\mathbf{s}[m]\bigg)\frac{\vec{\mathbf{r}}[n]-\vec{\mathbf{r}}[m]}{|\;\vec{\mathbf{r}}[n]-\vec{\mathbf{r}}[m]\;|^2} \\
&\propto \sum_n \bigg(\mathbf{s}[n]-\mathbf{s}[m]\bigg) \hat{\mathbf{r}}_{nm},
\end{aligned}
\end{equation}
where $\vec{\mathbf{g}}$ is a vector containing the derivatives at each source pixel, $\vec{\mathbf{r}}$ is a vector containing the position vectors of each source pixel, $\hat{\mathbf{r}}_{nm}$ is a unit vector pointing from $n$ to $m$, and the sums are over the four nearest pixels.
The last proportionality holds because, for a Cartesian grid, the distances between adjacent pixels are identical and can be absorbed into the regularisation strength.
To approximate the second derivative across the source plane, we write down the Laplacian as
\begin{equation}
\label{seconddereq}
\begin{aligned}
\mathbf{h}[m] &= \vec{\nabla} \cdot \vec{\mathbf{g}}[m] \\
&\propto \vec{\nabla} \cdot \sum_n \bigg(\mathbf{s}[n]-\mathbf{s}[m]\bigg) \hat{\mathbf{r}}_{nm} \\
&\propto \bigg(\sum_n \mathbf{s}[n] \bigg) -N\,\mathbf{s}[m],
\end{aligned}
\end{equation}
where $\mathbf{h}$ is a vector containing the second derivative at each source pixel and the sum is again over the $N=4$ nearest pixels.
In both cases, if the pixel is not on the edge of the grid then the sums include the four pixels to the immediate left, right, top, and bottom of the pixel in question.
If the pixel is on the edge, the ``missing'' pixels are assumed to contain zero flux.
This effectively assumes the surface brightness outside the source grid is zero, which can lead to ineffective regularisation if the grid is small enough that the edges are close to the region of interest.
\citet{suyureg} note that the derivative calculations can be modified to avoid assuming zero surface brightness outside the grid, but that can lead to problems for ranking lens models.\footnote{
As \citet{suyureg} explain, dropping the zero surface brightness assumption causes the Hessian of $\mathbf{R}$ to become singular.
The singularity can be removed by introducing a renormalisation constant, but the constant will vary with the lens model, complicating the model comparison.
}

\begin{figure}
\centering
\includegraphics[width=0.45\textwidth]{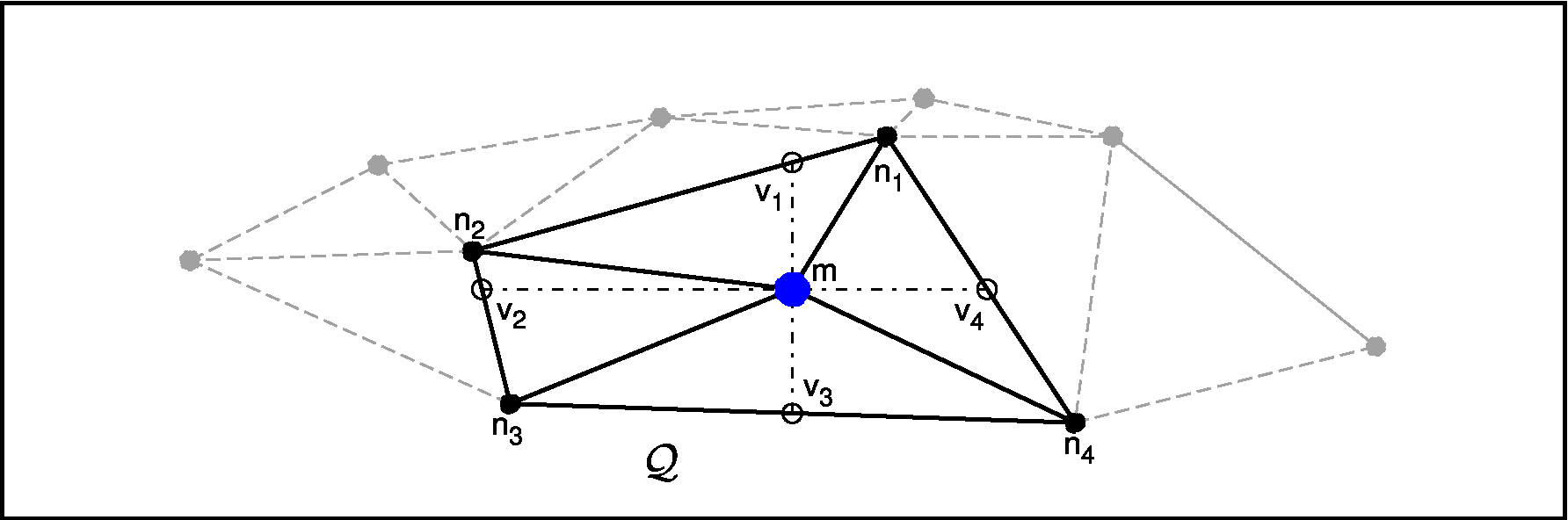}
\caption{\small
Diagram illustrating the derivative calculation using the FDM on an irregular grid.
The blue point indicates the pixel $m$ where we seek to compute the derivative.
The grid is the same as that shown in Fig.~\ref{fig:vegettigridmaking}.
The black, solid lines form a quadrilateral $\mathcal{Q}$ connecting the surrounding pixels (labeled $n_p$, where $p=\{1,2,3,4\}$).
The points where $\mathcal{Q}$ intersects the horizontal and vertical lines through $m$ are called virtual pixels (labeled $v_p$, where $p=\{1,2,3,4\}$).
The flux at each virtual pixel is a linear combination of the fluxes at the two surrounding pixels that are colinear with that virtual pixel.
The virtual pixels are used to compute the derivative at $m$.}
\label{fig:surrvirt}
\end{figure}

The preceding discussion can be extended to adaptive grids, although some care is needed because there may be more than four pixels nearby and it may not be immediately obvious which ones should be used.
\citet{vegettigrid} compute the derivative for a particular pixel $m$ using the triangles that surround $m$ in the Delaunay triangulation of the grid.
In \textit{pixsrc}, we instead identify four pixels (hereafter referred to as surrounding pixels) as follows.
Transforming to a coordinate system centered on $m$, we select the surrounding pixels so that each pixel lies in a different quadrant, each pixel is near $m$, and the quadrilateral $\mathcal{Q}$ formed by the pixels deviates the least from a square.
We use the surrounding pixels to calculate the surface brightness at the intersections of the $x$ and $y$ axes with $\mathcal{Q}$, which we refer to as virtual pixels.
(A schematic diagram of the surrounding and virtual pixels is shown in Fig.~\ref{fig:surrvirt}.)
From the surface brightnesses at the virtual pixels, we can compute the derivatives using Eq.~\ref{firstdereq} and a modified version of Eq.~\ref{seconddereq}.
After some algebra, the first derivative at $m$ can be expressed as
\begin{equation}
\begin{aligned}
\vec{\mathbf{g}}[m] = &\sum_n\bigg( \frac{D[v_{n-1},s_{n-1}]}{D[v_{n-1},m]D[s_{n-1},s_{n}]} \, \hat{\mathbf{r}}_{v_{n-1}}
\\
&\qquad + \frac{D[v_{n},s_{n+1}]}{D[v_{n},m]D[s_{n+1},s_{n}]} \, \hat{\mathbf{r}}_{v_{n}} \bigg)\mathbf{s}[s_n]
\\
&- \bigg( \sum_n\frac{1}{D[m,v_{n}]} \, \hat{\mathbf{r}}_{v_{n}} \bigg) \mathbf{s}[m],
\end{aligned}
\end{equation}
and the second derivative at $m$ is given by
\begin{equation}
\begin{aligned}
\mathbf{h}[m] = &\sum_n\bigg( \frac{D[v_{n-1},s_{n-1}]}{D[v_{n-1},m]D[s_{n-1},s_{n}]}
\\
&\qquad + \frac{D[v_{n},s_{n+1}]}{D[v_{n},m]D[s_{n+1},s_{n}]} \bigg)\mathbf{s}[s_n]
\\
& - \bigg( \sum_n\frac{1}{D[m,v_{n}]} \bigg) \mathbf{s}[m],
\end{aligned}
\end{equation}
where the sums run from $n=1$ to $n=4$, $D[r,s]$ is a functional that returns the distance between points $r$ and $s$, $v_p$ is the $p^\text{th}$ virtual pixel, and $\hat{\mathbf{r}}_{v_p}$ is a unit vector pointing toward the $p^\text{th}$ virtual pixel.
For simplicity of notation, we let the indices wrap around (e.g., $v_0=v_4$ and $v_5=v_1$).

Fig.~\ref{fig:reg0} suggests that derivatives calculated with the FDM can be inaccurate.
Certain configurations of points on the fully adaptive grid can cause virtual pixels to lie very close to $m$, leading to an anomalously high estimate for the derivative.
Such events are rare and do not have a dramatic effect on the source reconstruction.
The adaptive Cartesian grid is less susceptible to such gridding issues.

\subsubsection{Divergence theorem method}
\label{sec:dtm}

\begin{figure}
\centering
\includegraphics[width=0.45\textwidth]{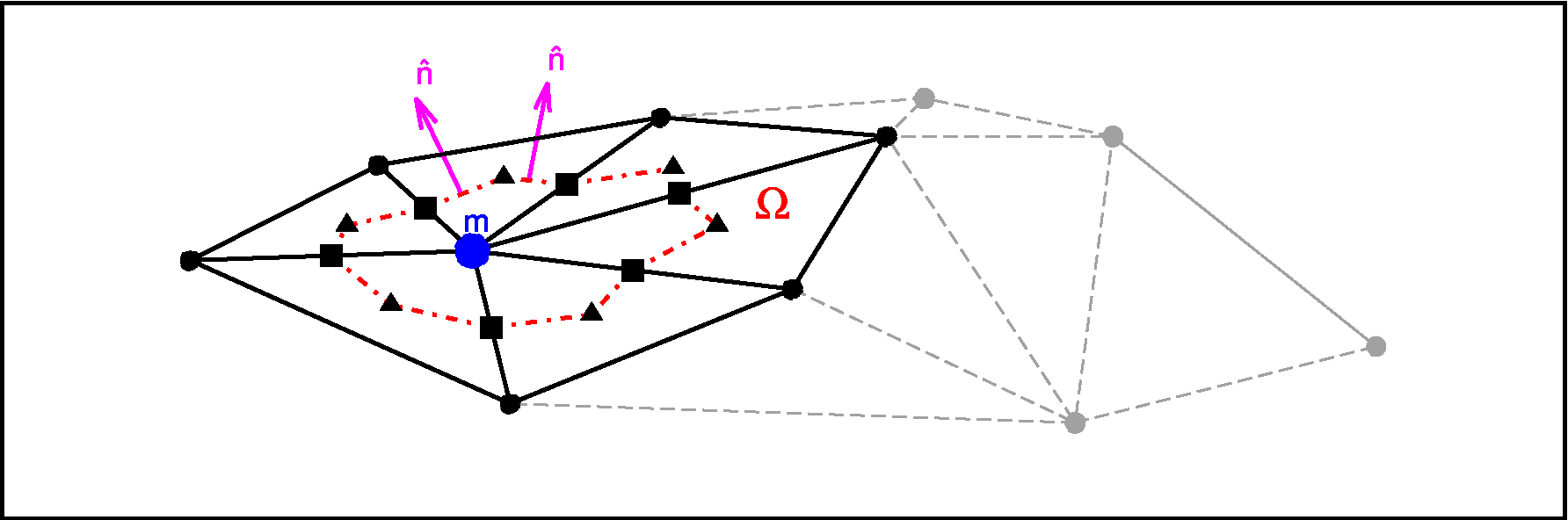}
\caption{\small
Diagram illustrating the derivative calculation using the DTM.
The blue point again indicates the pixel $m$ where we seek to compute the derivative.
The black, solid lines form a polygon surrounding $m$.
The square points are placed at the midpoints between $m$ and the surrounding pixels.
The triangle points are placed at the centroids of the triangles.
The stencil $\Omega$ is formed by connected midpoints to adjacent centroids.
The unit vector $\hat{n}$, denoted by the magenta arrows, is orthogonal to the edges of the stencil, points outwards, and changes direction as the stencil is traced along its edges.
This figure is inspired by Fig.~1 in \citet{citeDTM}.}
\label{fig:dtmstencil}
\end{figure}

The method described here is developed in \citet{citeDTM}; we reproduce some of the key elements.
It is called by the authors an irregular grid finite difference method based on the Green-Gauss theorem (as Green's theorem reduces to Gauss' theorem in two dimensions).
The theorem states that for a scalar function $F$ defined on $\mathbb{R}^2$ with continuous partial derivatives, we can relate the surface integral over some region $\Omega$ to a line integral along the boundary of $\Omega$:
\begin{equation}
\int \int_\Omega \vec{\nabla}F\,\mathrm{d}\Omega = \oint_{\partial\Omega} F\hat{n}\,\mathrm{d}s,
\end{equation}
where $\hat{n}$ is a unit normal vector on the boundary, pointing outwards.
Suppose $\Omega$ is a region, called a stencil, small enough that $\vec{\nabla}F$ is approximately constant across the region.
Then we can pull the gradient out of the integral and write
\begin{equation}
\label{greengauss1eq}
  \vec{\nabla}F = \frac{1}{\Omega}\oint_{\partial\Omega} F\hat{n}\,\mathrm{d}s,
\end{equation}
from which it follows that 
\begin{equation}
\label{greengauss2eq}
  \vec{\nabla}\cdot(\vec{\nabla}F) = \frac{1}{\Omega}\oint_{\partial\Omega} \hat{n}\cdot\vec{\nabla} F\,\mathrm{d}s,
\end{equation}
For implementation, the integrals are converted to sums, and $\Omega$ is defined by connecting centroids of Delaunay triangles to adjacent midpoints of the sides of the triangles (see Fig.~\ref{fig:dtmstencil}).
Depending on the density of source pixels, the stencil $\Omega$ may not be small enough for $\nabla F$ to be constant.
We nevertheless take Eq.~\ref{greengauss1eq} to define an effective gradient for each pixel.\footnote{Implementation of the second derivative requires additional correction terms found in \citet{citeDTM}; it is still under refinement in \textit{pixsrc}.}

Unlike the FDM, the DTM does not assume the flux vanishes outside the grid.
Eq.~\ref{greengauss1eq} and \ref{greengauss2eq} can be applied to the grid edges, as long as care is taken in closing the line integrals.
Thus, edge effects in the regularisation are minimal.
Fig.~\ref{fig:reg0} suggests that derivatives computed with the DTM are more accurate than those computed with the FDM, because the DTM uses all nearby pixels.

\subsubsection{Analytic source regularisation}

As an alternative to derivative-based regularisation, we have developed a quadratic form of regularisation that penalises the source for deviations in surface brightness from one or more analytic profiles.
We find that analytic source regularisation (ASR) is especially useful in recovering the surface brightness of the source in noisy data.
Currently, the reference surface brightness distribution has a S\'{e}rsic profile,
\begin{equation}
I(\vec{r}) = I_0 \exp{\Bigg[-\bigg(\frac{|\vec{r}|}{r_\text{s}}\bigg)^{1/n}\Bigg]},
\end{equation}
where $I_0$ is the normalisation, $r_\text{s}$ is the scale radius, and $n$ is the S\'{e}rsic index.
Elliptical models are created from a linear transformation of coordinates.
More complicated sources can be built from a combination of S\'{e}rsic profiles that represent multiple, blended sources (such as ``knots'' in star-forming galaxies).

\begin{figure}
\centering
\includegraphics[width=0.45\textwidth]{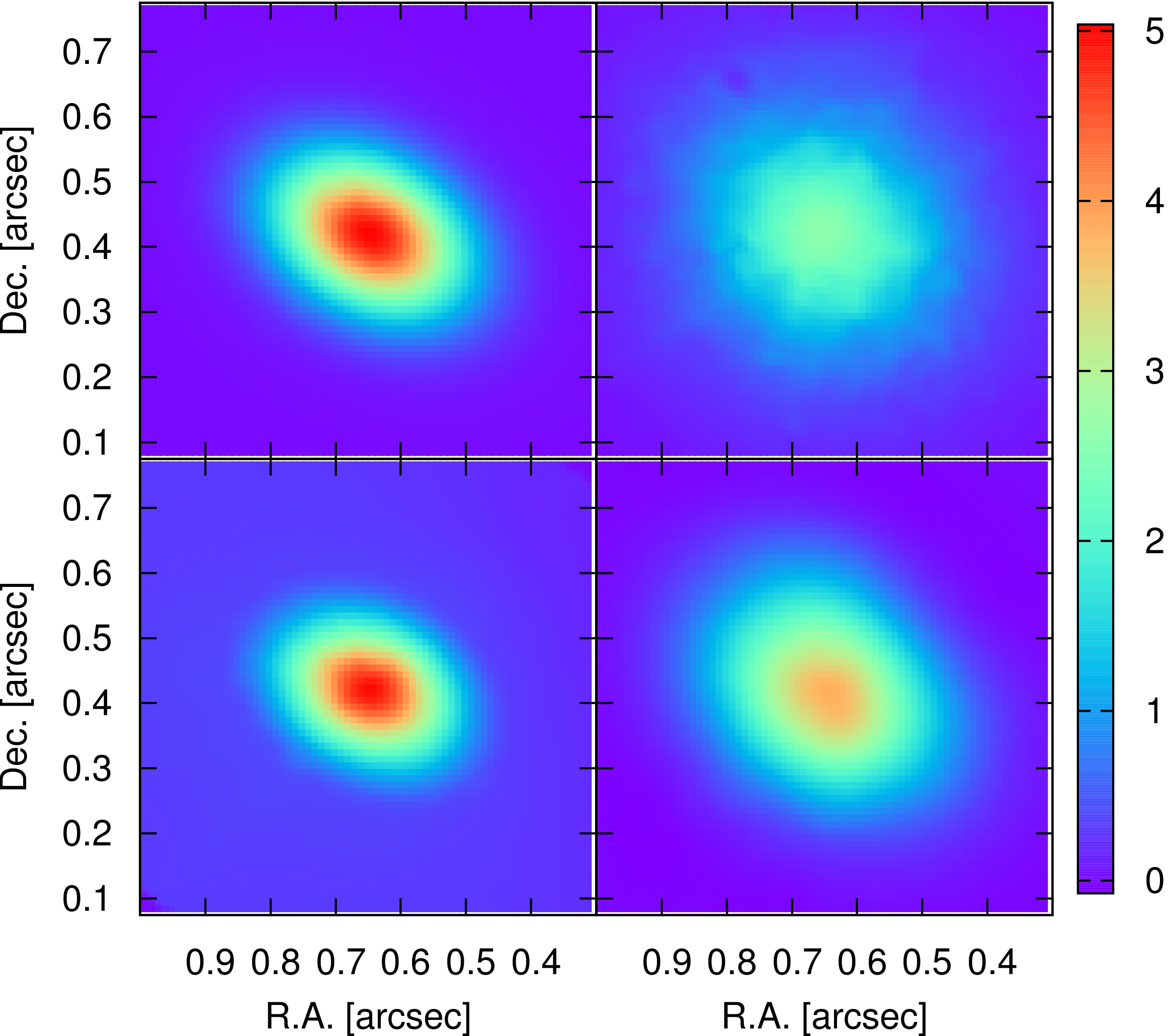}
\caption{\small
Comparison of sources reconstructed from noisy data.
The test data have a source in the cusp configuration with a peak S/N of 1 (see the top left panel of Fig.~\ref{fig:mockdatanoise}) and a resolution of 0.05 arcsec/pixel.
Clockwise from top left: the true source surface brightness followed by sources reconstructed from gradient-based regularisation, curvature-based regularisation, and analytic source regularisation.
The source recovered using ASR best matches the true source brightness.
In the case of blended sources (not shown), ASR also outperforms derivative-based regularisation.
}
\label{fig:asr}
\end{figure}

To implement ASR, we first find the analytic source $\mathbf{s}_a$ that best fits the data.
We vary the position, normalisation, scale radius, S\'{e}rsic index, ellipticity, and position angle of the analytic source using a downhill simplex optimisation routine \citep{numericalrecipes}.
We then use the best-fit analytic source to construct a regularisation matrix, $\mathbf{H}$, that acts on a source vector, $\mathbf{s}$, to produce a deviation vector, $\mathbf{\Delta}=\mathbf{Hs}$, \CRKadd{whose value} at pixel $m$ is given by
\begin{equation}
\label{eq:deltaasr}
\mathbf{\delta}[m] = \bigg( \sum_n \frac{\mathbf{s}[n]}{\mathbf{s}_a[n]}  \bigg) -N \frac{\mathbf{s}[m]}{\mathbf{s}_a[m]},
\end{equation}
where $\mathbf{s}_a[p]$ is the flux at $p$ from the analytic source, and the sum is over $N$ \CRKadd{pixels that share a Delaunay triangle with pixel $m$.
The deviation vector vanishes if the source vector agrees completely with the analytic profile, or indeed if $\mathbf{s}$ is any real multiple of $\mathbf{s}_a$.}\footnote{
Because ASR can obtain a minimum for $\mathbf{s}\neq\mathbf{0}$, some of the algebra in \S\ref{sec:bayesian} is modified.
However, the key results (specifically Eqs.~\ref{eq:smp}--\ref{eq:lasteqbayesian}) are unchanged.
}
\CRKadd{More generally, $\mathbf{\Delta}$ quantifies the degree to which $\mathbf{s}$ does not match a multiple of $\mathbf{s}_a$.}
Because the inverse brightness values in Eq.~\ref{eq:deltaasr} can become large toward the outer regions of the analytic profile, we set the analytic source flux to 10\% of the noise level once it falls below this value.
Fig.~\ref{fig:asr} suggests that ASR is more effective than derivative-based regularisation at recovering the source from noisy data.
This result is perhaps not surprising; because ASR assumes a functional form for the source, it is a stronger prior than derivative-based regularisation.
It is important to note that ASR will yield accurate source reconstructions only if the true source is well described by the assumed functional form.

\begin{figure}
  \centering
  \includegraphics[width=0.45\textwidth]{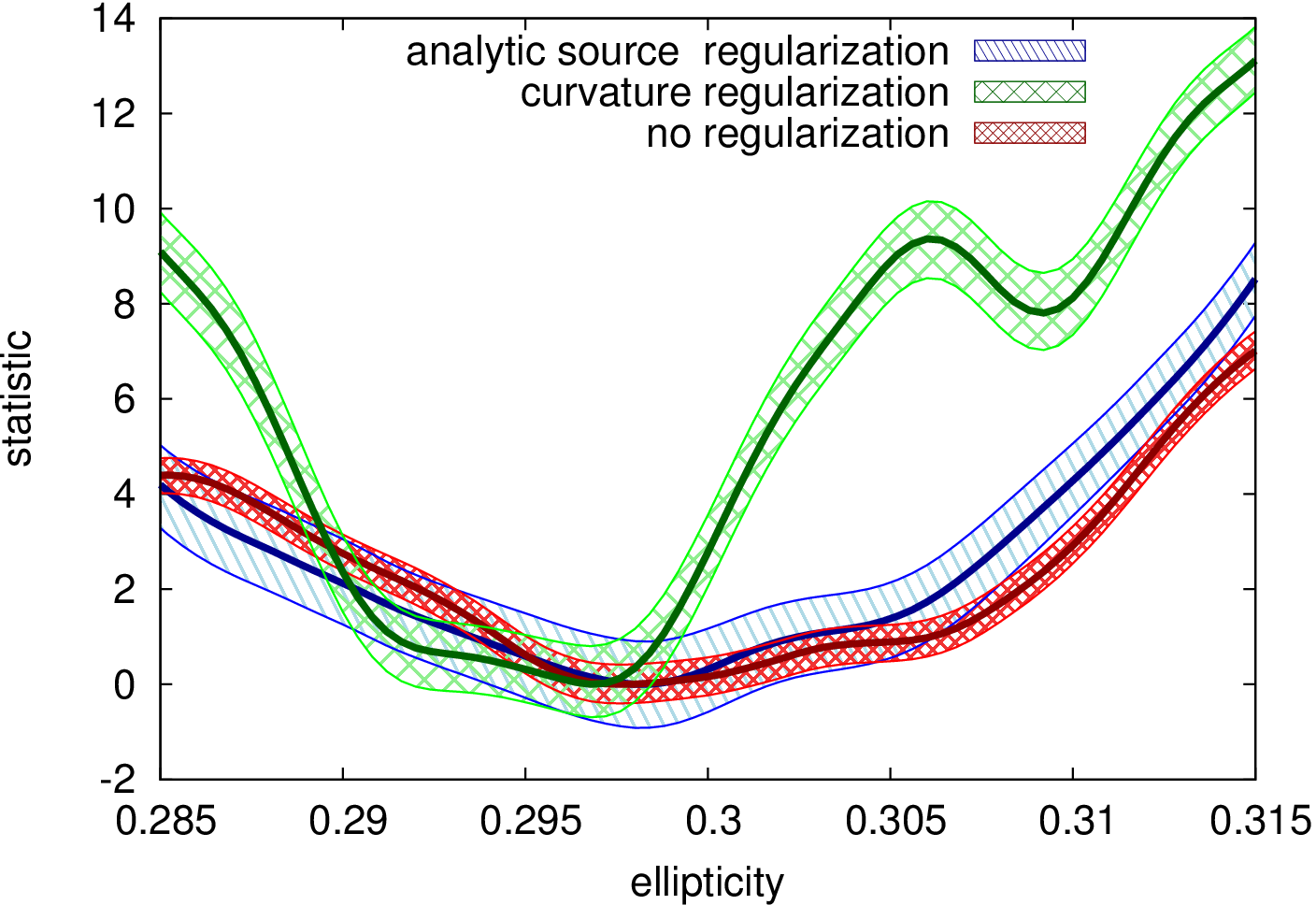}
  \caption{\small
Effects of regularisation on parameter estimation.
The $y$-axis label ``statistic'' denotes $\chi^2$ and $-2\ln \mathcal{E}$ for cases without and with regularisation, respectively, but for simplicity we refer to both as $\chi^2$.
(We have applied a vertical offset to facilitate comparing the curves.)
We construct many realisations of a cusp lens with a peak S/N of unity and a resolution of 0.1 arcsec/pixel.
After stacking the results, we expect the $\chi^2$ curve without regularisation (shown in red) to represent the true errors on the ellipticity.
ASR (shown in blue) seems to agree well with the reference case.
By contrast, curvature regularisation (shown in green) has a minimum that is shifted away from the true value $e=0.3$.
Also, the $\chi^2$ curve rises rapidly away from the minimum, causing the parameter uncertainties to be underestimated.
For each case, the dark, thick line corresponds to the median value, and the bands are 68\% confidence intervals, estimated from bootstrapping.
}
\label{fig:regvsnoreg}
\end{figure}

At this point we should consider whether regularisation introduces any biases in the values or uncertainties for recovered lens model parameters.
Because the noise is Gaussian and centered on zero, we conjecture that analysing many different realisations of the noise can uncover the true underlying likelihood function (as an alternative to explicitly regularising the source surface brightness).
We construct thousands of ``observations'' of a cusp lens with a peak S/N of unity and a resolution of 0.1 arcsec/pixel.
We vary the ellipticity of the lens while holding other parameters fixed at their true values.
The $\chi^2$ curves from individual runs vary significantly, but stacking the results washes away the fluctuations from noise (see the red curve in Fig.~\ref{fig:regvsnoreg}).
The stacked curve from ASR (shown in blue) matches the underlying $\chi^2$ curve well.
The results from curvature regularisation, by contrast, show a small bias toward lower ellipticity and underestimate the uncertainties for this parameter.
This is yet another indication that ASR can outperform derivative-based regularisation when the data are noisy and the true source follows an analytic profile.
At higher S/N (not shown), there is less difference between the regularisation schemes.

%%%%%%%%%%%%%%%%%%%%%%%%%%%%%%%%%%%%%%%%%%%%%%%%%%%%%%%%%%%%
\subsection{Effect on $\chi^2$}
\label{sec:obstacles}

When exploring the lens model parameter space, we find that the likelihood surface can be jagged even for our clean test data.
\citet{citewallingtonMEM} remarked on ``glitches'' in $\chi^2$ for their maximum entropy analysis, but noted that the glitches disappeared as the PSF and noise vanished.
In our analysis, the jaggedness is reduced but not eliminated in that limit.
It arises, we suspect, from the discrete nature of PBSR itself.
A small, continuous change in the lens model parameters can shift the source pixels in a way that causes the Delaunay algorithm to connect the pixels in a different way, leading to abrupt changes in the lensing operator and regularisation matrix.

\begin{figure*}
\centering
  \begin{tabular}{@{}c@{\hskip 0.1\textwidth}c@{}}
    \includegraphics[width=0.4\textwidth]{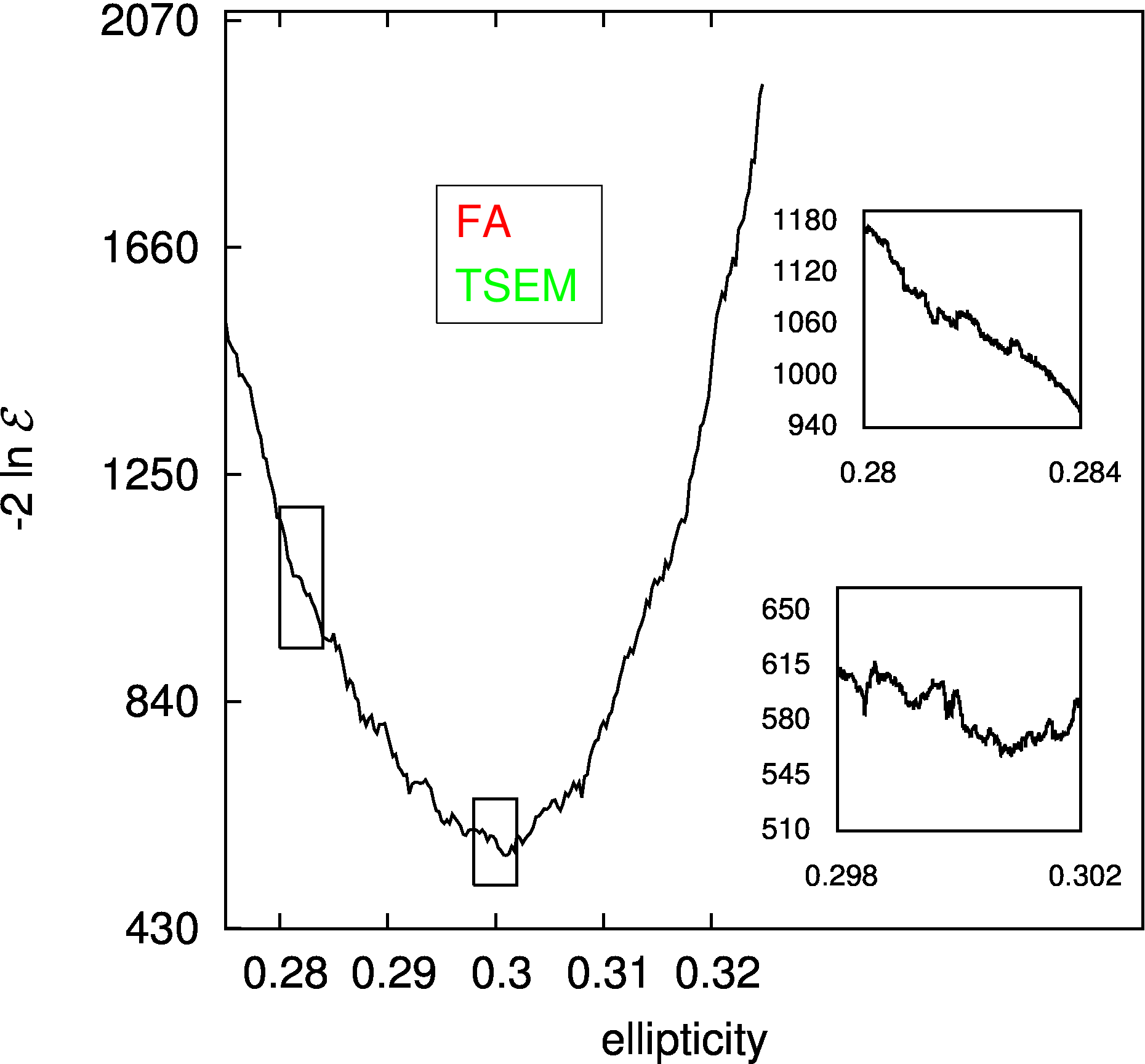} &
    \includegraphics[width=0.4\textwidth]{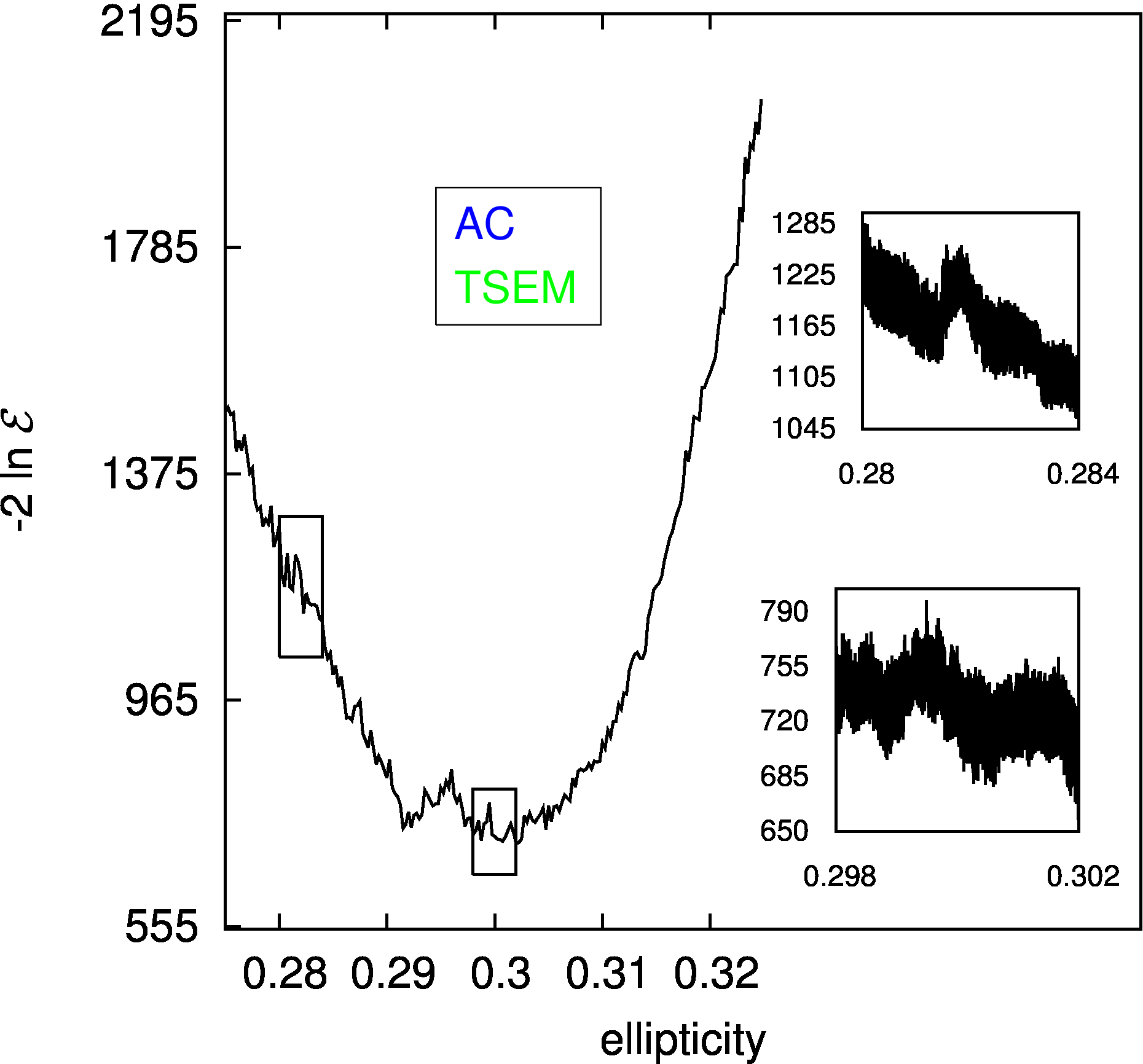} \\
    \includegraphics[width=0.4\textwidth]{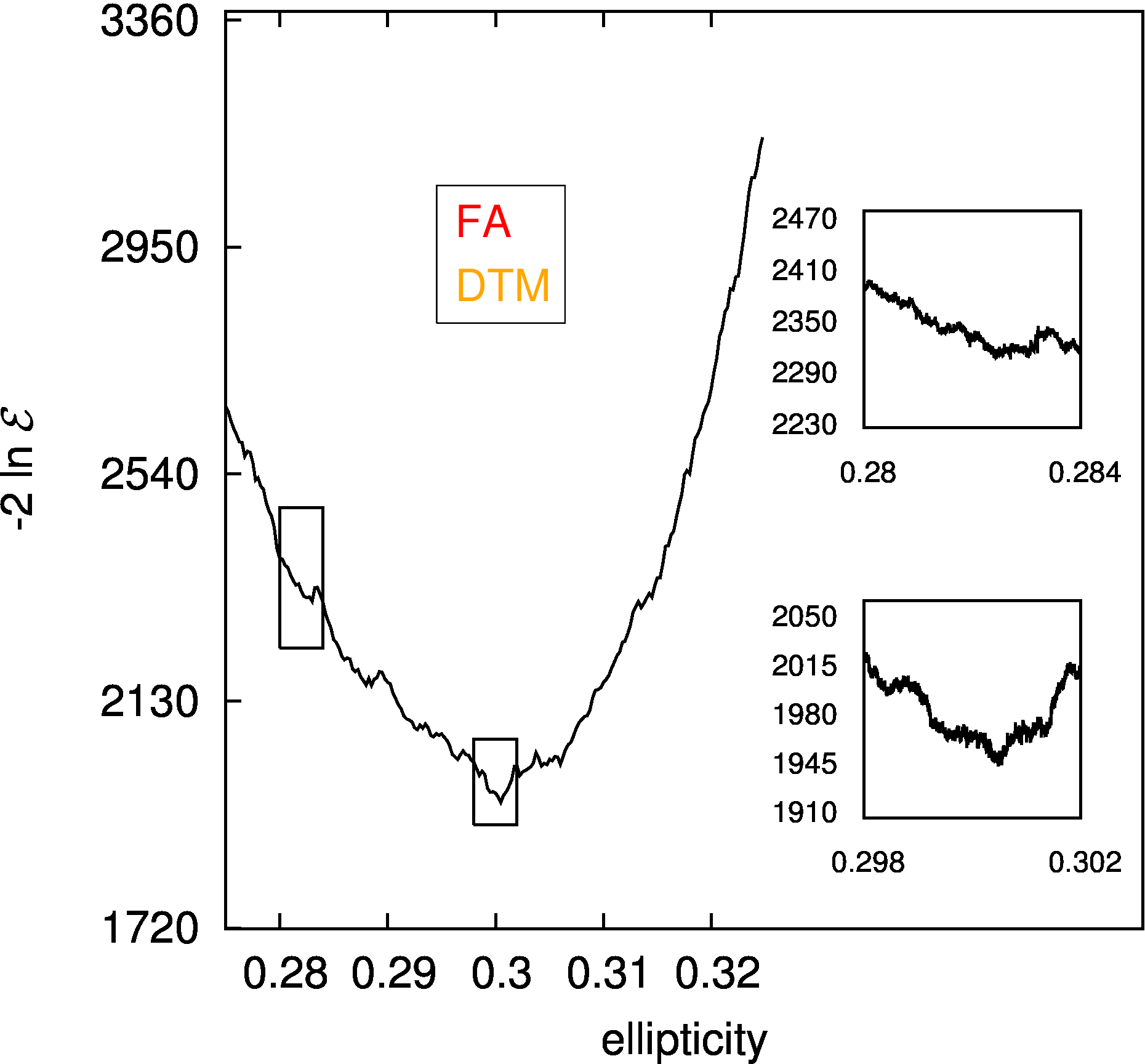} &
    \includegraphics[width=0.4\textwidth]{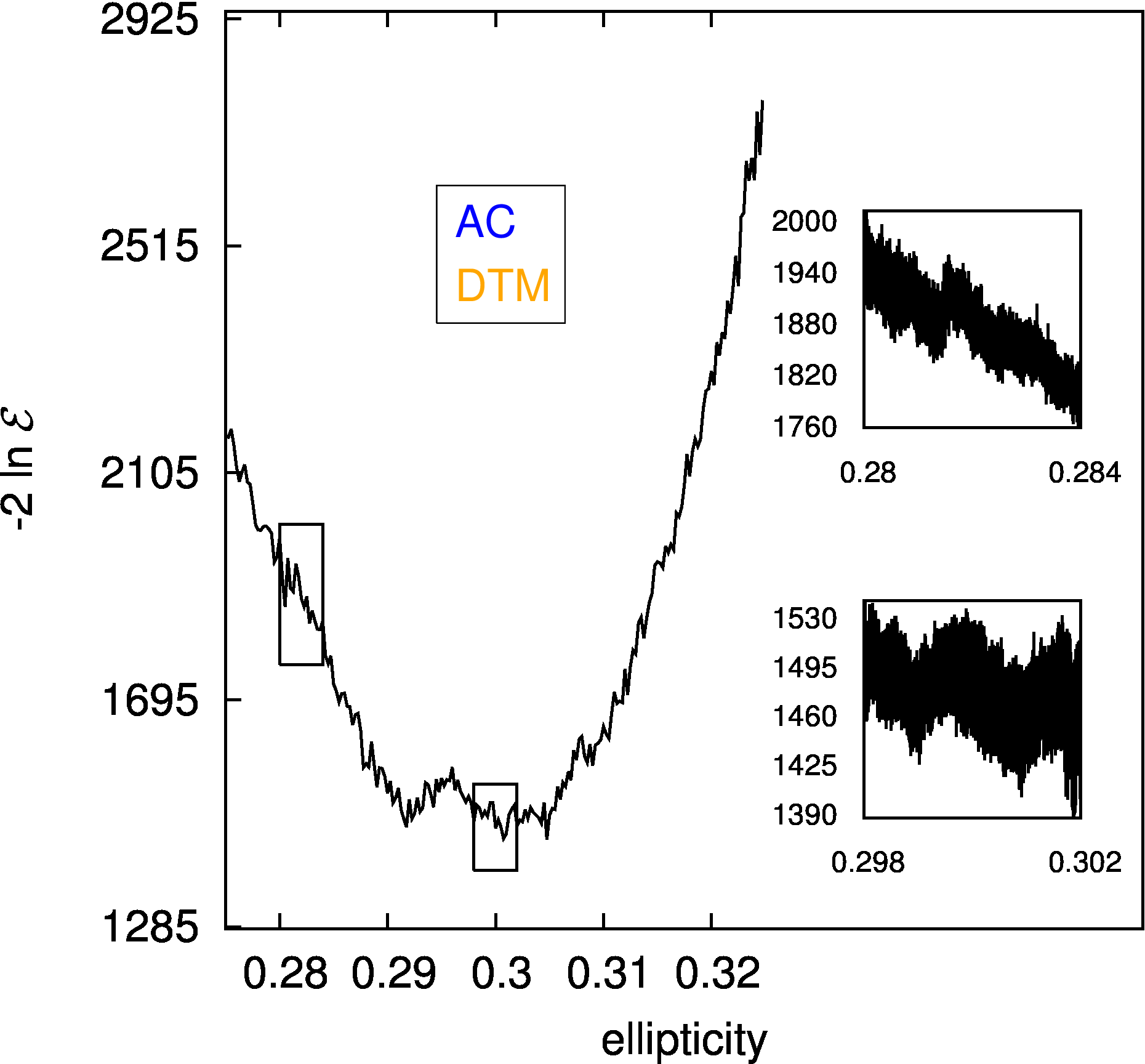} \\
    \includegraphics[width=0.4\textwidth]{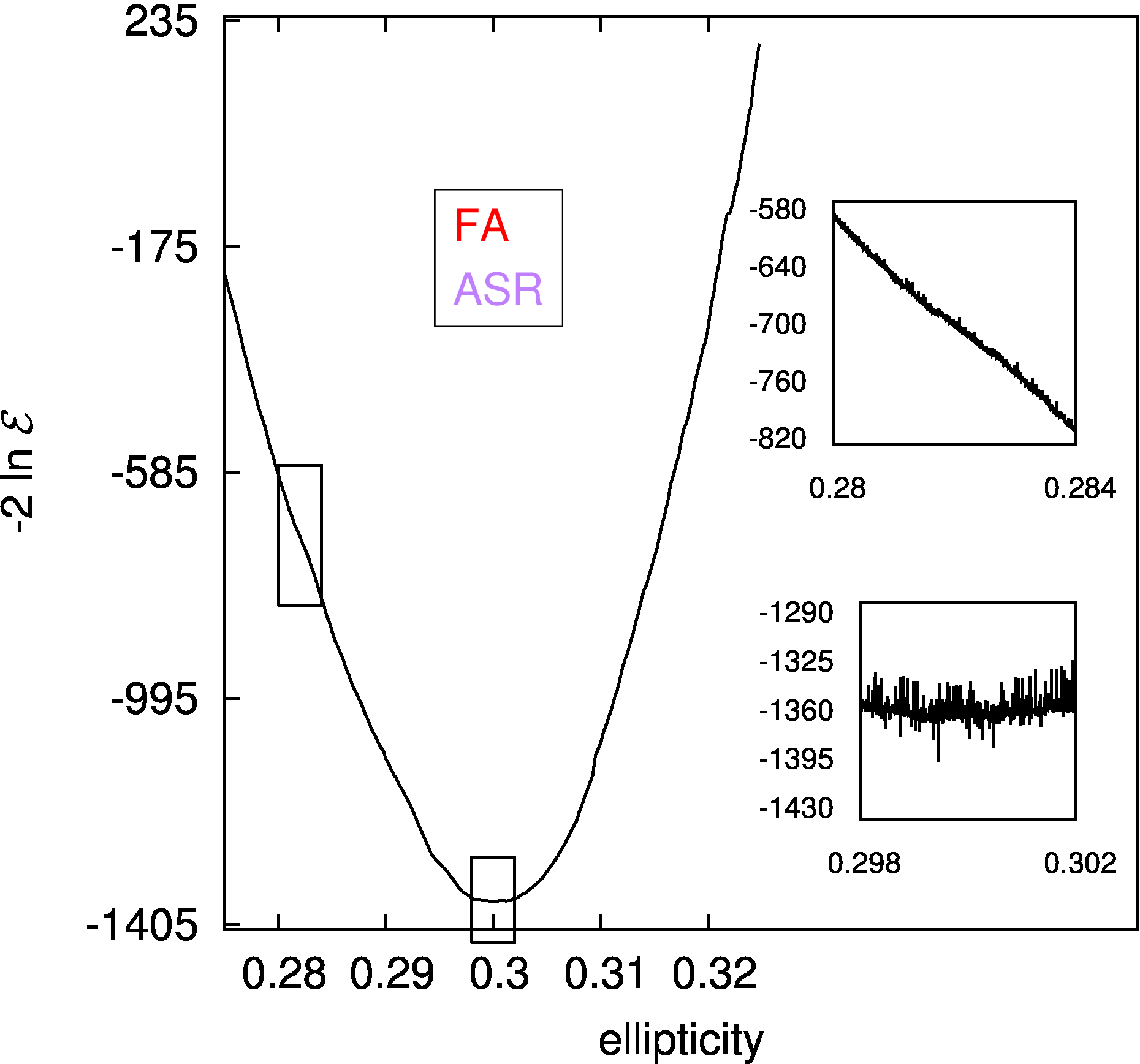} &
    \includegraphics[width=0.4\textwidth]{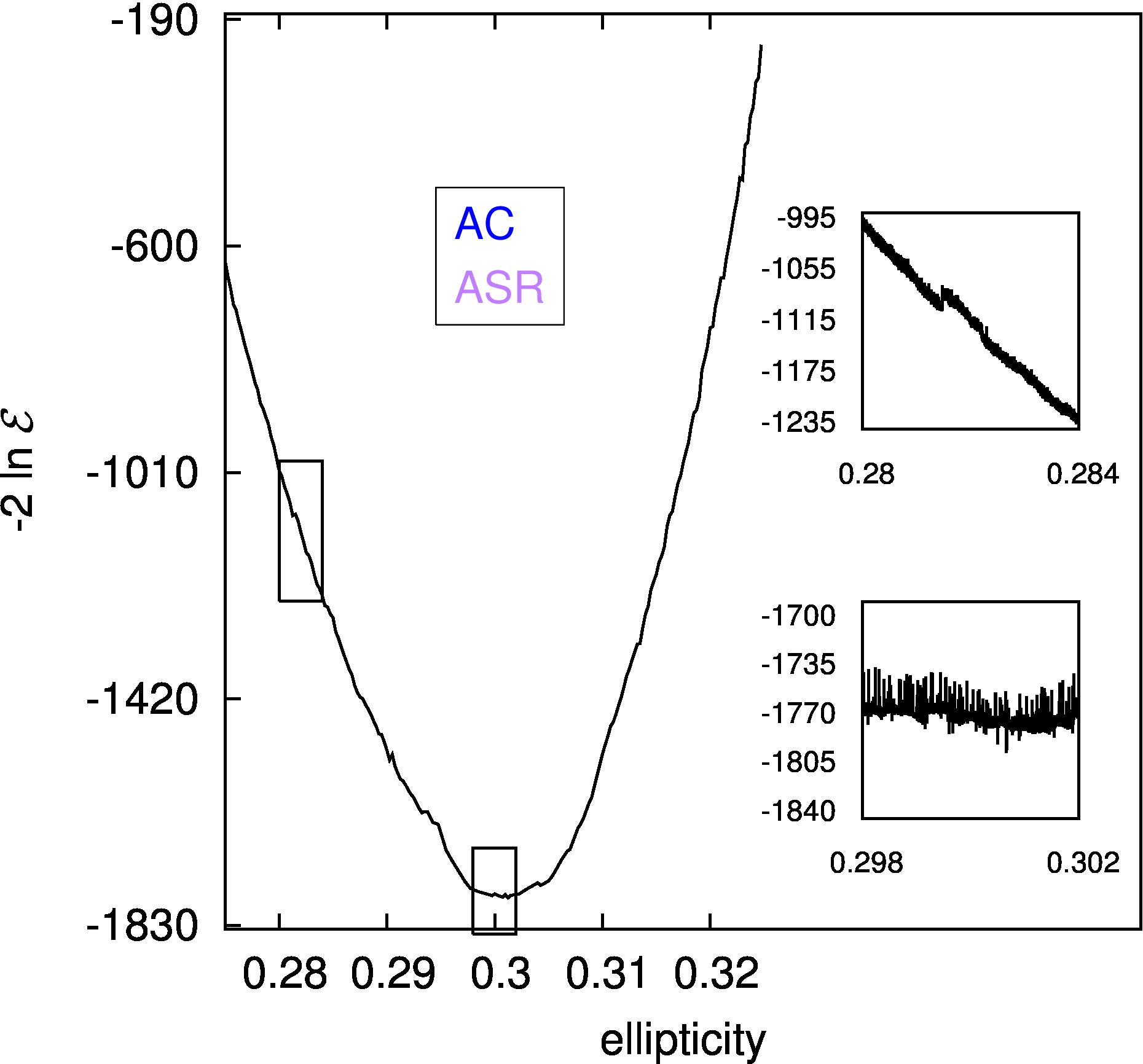}
  \end{tabular}
  \caption{\small
Effective $\chi^2$ as a function of ellipticity for various gridding and regularisation schemes.
The test data have a source in the cusp configuration with a peak S/N of 25 and a resolution of 0.05 arcsec/pixel.
The columns correspond to different grids (left is fully adaptive, right is adaptive Cartesian).
The rows correspond to different regularisation schemes (top is gradient regularisation with FDM, middle is gradient regularisation with DTM, bottom is analytic source regularisation [ASR]).
\CRKadd{The different panels have the same vertical range (and the vertical offsets are not meaningful).}
In general, the fully adaptive grid leads to less noise in $\chi^2$ than the adaptive Cartesian grid.
ASR produces the smoothest curve over large scales, presumably because the regularisation matrix does not change discretely and the deviation from an analytic profile is measured in a dimensionless way.
}
\label{fig:1devicut}
\end{figure*}

To probe these issues, we examine one-dimensional cuts of the $\chi^2$ surface for various gridding and regularisation schemes.
We focus on test data for the cusp configuration with a peak S/N of 25 and pixel scale of $0.05$ arcsec/pixel.
We fix all lens model parameters at their correct values and vary only the ellipticity of the lensing galaxy.
We consider different combinations of grids (fully adaptive or adaptive Cartesian) and priors (gradient regularisation with FDM or DTM, or ASR).
The results are shown in Fig.~\ref{fig:1devicut}.

Qualitatively, we find that the fully adaptive grid shows less small-scale fluctuation in $\chi^2$ than the adaptive Cartesian grid.
Since the fully adaptive grid is constructed by ray tracing image pixels to the source plane, it more naturally accommodates small changes in the lens model.
The adaptive Cartesian grid, by contrast, either remains fixed or changes discretely (if the magnification crosses the criterion for subgridding; see \S\ref{sec:adacargri}).
Thus, even though the adaptive Cartesian grid yields more accurate derivative calculations (recall Fig.~\ref{fig:reg0}), that benefit seems to be outweighed by gridding noise in $\chi^2$.
It may be possible to improve the performance of the adaptive Cartesian grid by developing a different criterion for subgridding, but such modifications have not yet been explored.

Turning to regularisation, the DTM yields somewhat smaller fluctuations than the FDM, at least for the fully adaptive grid (with the adaptive Cartesian grid, the noise is dominated by the gridding anyway).
ASR leads to the smoothest $\chi^2$ curves for both types of grids.
As the lens model parameters vary, the best fit analytic source and the corresponding weights in the regularisation matrix can vary smoothly as well.
It is interesting that the fully adaptive grid with the DTM does not show a similar level of smoothness, because that method also changes continuously with lens model parameters.
The difference may occur because the deviation vector $\mathbf{\delta}$ in Eq.~\ref{eq:deltaasr} is a dimensionless ratio of surface brightnesses, whereas the derivatives used for gradient or curvature regularisation have units of surface brightness divided by distance or squared distance.
Using a dimensionless measure of deviation allows each pixel to have equal weight in the regularisation matrix, a quality that the derivative-based methods do not necessarily have.

Finally, we note that the $\chi^2$ curve is flatter near the minimum for ASR than it is for gradient regularisation (focusing now on the fully adaptive grid).
This causes ASR to yield larger uncertainties in the ellipticity of the lensing galaxy, as we saw already in Fig.~\ref{fig:regvsnoreg}.
If the ASR is taken to represent the true posterior probability distribution, then the errors reported using the fully adaptive grid with gradient regularisation are being underestimated.

In summary, we find that the gridding and regularisation schemes both affect the level of noise in the Bayesian evidence.
These two algorithmic issues need to be considered carefully in applications of PBSR.

%%%%%%%%%%%%%%%%%%%%%%%%%%%%%%%%%%%%%%%%%%%%%%%%%%%%%%%%%%%%
\section{Practical issues}
\label{sec:mockdatab}
%%%%%%%%%%%%%%%%%%%%%%%%%%%%%%%%%%%%%%%%%%%%%%%%%%%%%%%%%%%%

In real data, the image resolution is typically fixed by the observational equipment, but the telescope pointing and the noise in the data are particular realisations; on a different day, the same observation would not actually be identical.\footnote{
Observations often include multiple exposures to handle cosmic rays, bad pixels, dithering, and subsampling the PSF.
We imagine our analysis being applied to the final image after data reduction.
}
We now consider whether such chance events introduce any statistical or systematic uncertainties into conclusions derived from lens modeling.
We examine noise and pointing both separately and jointly, with and without a PSF,\footnote{The PSF is used both in creating the mock data and in modeling the lens.} sometimes just optimising the parameters and sometimes performing a full parameter space exploration.
We assume Gaussian noise with zero mean, which can be considered to represent electron read-out noise, Poisson noise (in the large mean limit), or sky noise.
As a fiducial case, we use a lens in the cusp configuration with a pixel scale of 0.05 arcsec/pixel and a peak S/N of 10, but we examine different choices as discussed below.
Since ASR is computationally expensive, and curvature regularisation is well suited for initial parameter space explorations (see \S\ref{sec:conclusions} for more discussion), we use the fully adaptive grid with curvature regularisation and FDM here.

%%%%%%%%%%%%%%%%%%%%%%%%%%%%%%%%%%%%%%%%%%%%%%%%%%%%%%%%%%%%
\subsection{Effects of noise}
\label{sec:noise}

While the noise level will affect the uncertainty in lens model parameters, the particular noise realisation will also affect the best-fit values of the parameters.
To explore this possibility, we create 100 ``observations'' with the same data but different realisations of the noise, for the various noise levels shown in Fig.~\ref{fig:mockdatanoise}.
We optimise the parameters and examine the scatter among best-fit values (at this point we are not fully quantifying the parameter uncertainties).
The pixel scale is fixed at the high resolution of 0.02 arcsec/pixel (see Fig.~\ref{fig:mockdatares}) so that effects due to pixel size are minimised.

\begin{figure*}
\centering
\includegraphics[width=0.8\textwidth]{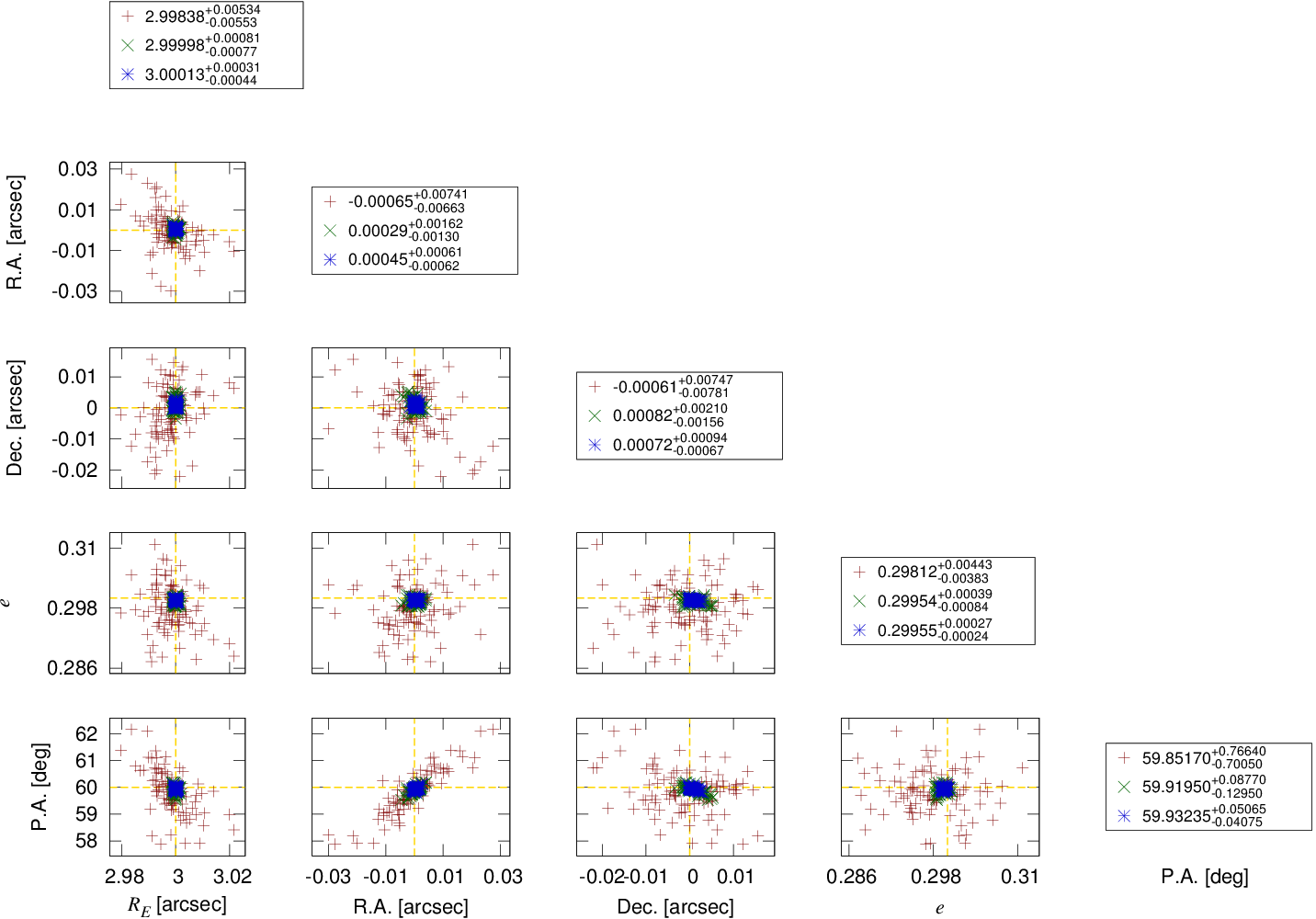}
\caption{\small
Best-fit lens model parameters for different realisation of noise in the data.
The source is in the cusp configuration.
Red, green, and blue points correspond to peak S/N levels of 1, 10, and 25, respectively.
The points marked correspond to optimal lens model parameters; this analysis does not include full parameter uncertainties.
The quoted uncertainties indicate the ranges that enclose 68\% of the best-fit values.
The pixel scale is fixed at 0.02 arcsec/pixel so the effects of pixel size are minimal.
Dashed, yellow lines mark the true values.}
\label{fig:noise_fa}
\end{figure*}

Fig.~\ref{fig:noise_fa} shows the results in terms of different two-dimensional parameter projections, along with the median and 68\% confidence intervals for individual parameters.
There is no significant bias in the parameter values.
Empirically, the scatter among best-fit values appears to have a power law dependence on S/N with a slope of $\sim -0.8$ across all lens model parameters.

%%%%%%%%%%%%%%%%%%%%%%%%%%%%%%%%%%%%%%%%%%%%%%%%%%%%%%%%%%%%
\subsection{Effects of pointing}
\label{sec:dither}

Telescope pointing affects how photons are collected into pixels, so small shifts may influence the data and hence the recovered model parameters.
To explore this issue, we again create 100 ``observations'' in which the pointing is shifted randomly.
The shifts are drawn from a uniform distribution that is one pixel in each direction,\footnote{
Ignoring edge effects, shifts of $N + \Delta x$ are equivalent to shifts of $\Delta x$, where $N$ is an integer.
}
for image resolutions of 0.03, 0.05, and 0.1 arcsec/pixel.
\textit{pixsrc} requires some amount of noise, but the noise map is kept identical and the noise level is minimal (the peak S/N is $5\times10^5$) so the effects are negligible.

\begin{figure*}
\centering
\includegraphics[width=0.8\textwidth]{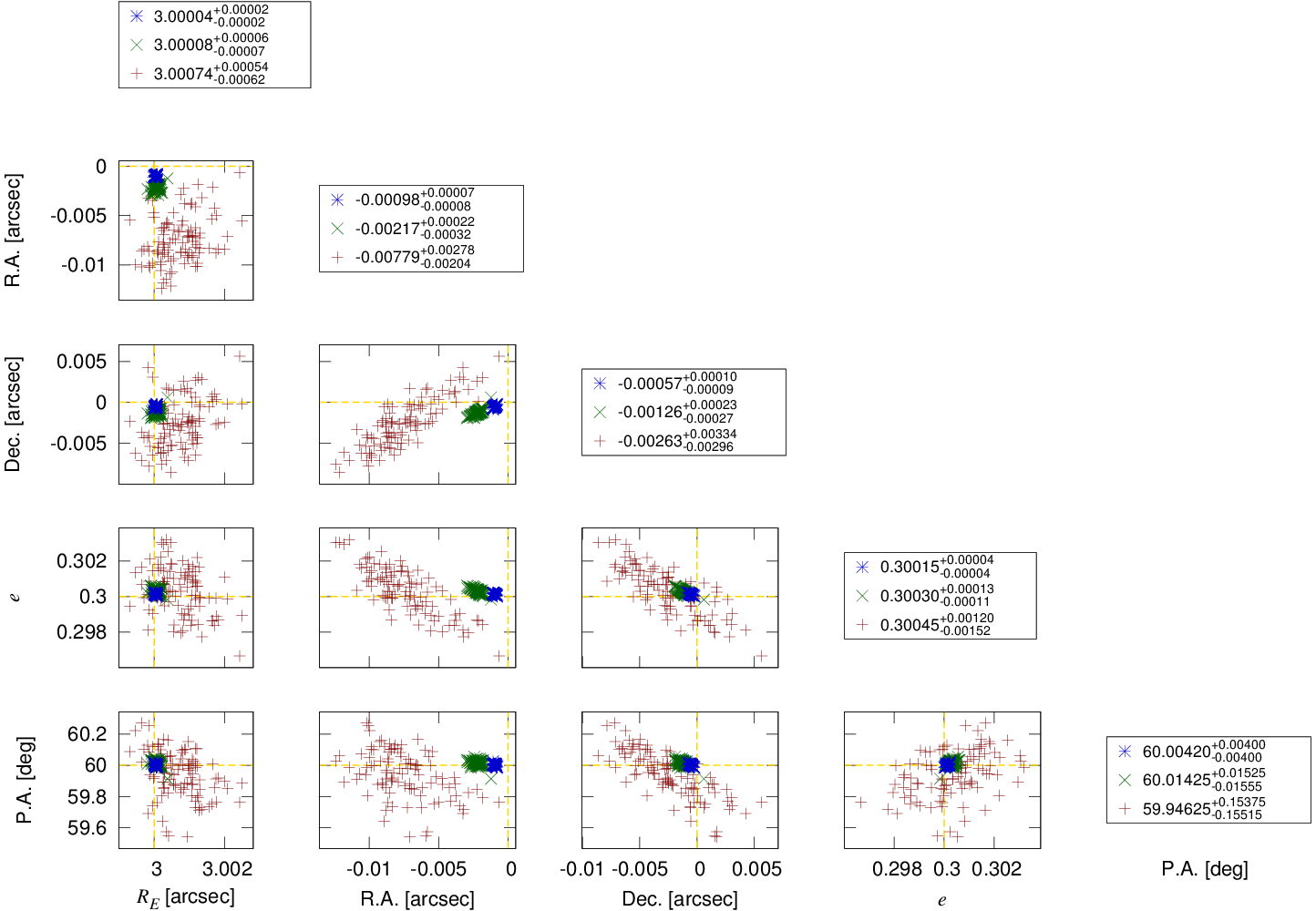}
\caption{\small
Best-fit lens model parameters for different realisations of the telescope pointing.
The source is in the cusp configuration.
The shifts are drawn from a uniform distribution that is one pixel in each direction.
Red, green, and blue points correspond to image resolutions of 0.1, 0.05, and 0.03 arcsec/pixel, respectively.
The peak S/N is $5\times10^5$ so that effects related to noise are negligible.
Dashed, yellow lines mark the true values.}
\label{fig:dither_fa}
\end{figure*}

\begin{figure*}
\centering
\includegraphics[width=0.8\textwidth]{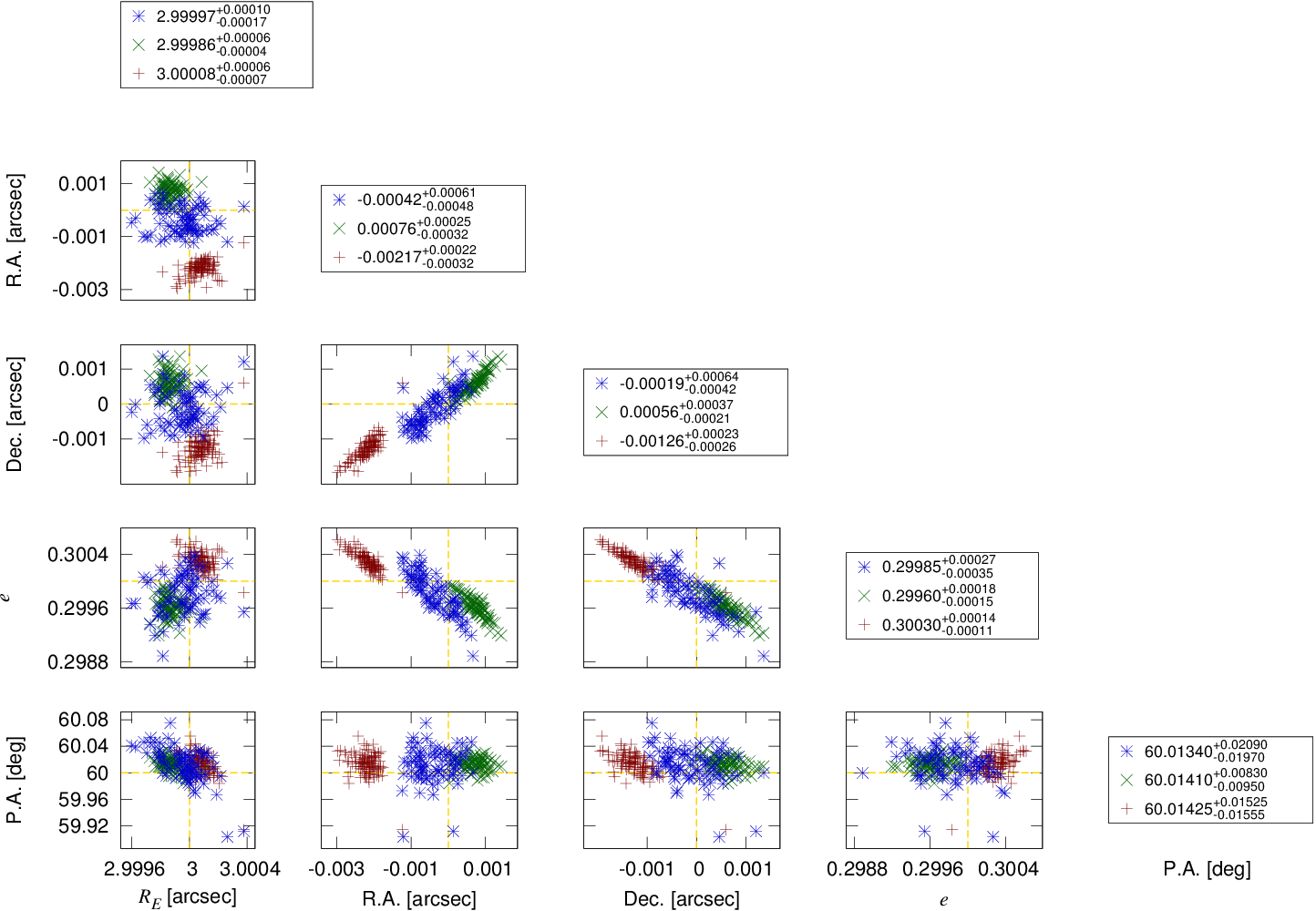}
\caption{\small
Similar to Fig.~\ref{fig:dither_fa} but including a PSF.
The source is placed in the cusp configuration, and the image resolution is fixed at 0.05 arcsec/pixel.
The red points have no PSF, while the green and blue points have circular Gaussian PSFs with FWHM equal to 0.059 and 0.12 arcsec, respectively.
}
\label{fig:ditherpsf}
\end{figure*}

Fig.~\ref{fig:dither_fa} shows two-dimensional projections of the best-fit parameter values.
The median values reveal biases that are small (a fraction of a pixel for the Einstein radius and position of the lens galaxy) but statistically significant.
The biases become less significant, however, when a PSF is included (see Fig.~\ref{fig:ditherpsf}).
For the case with no PSF, the scatter in the best-fit parameter values follows a power law with a slope of $\sim 3.3$ in terms of the linear pixel scale, and it increases further with the addition of a PSF.

%%%%%%%%%%%%%%%%%%%%%%%%%%%%%%%%%%%%%%%%%%%%%%%%%%%%%%%%%%%%
\subsection{Effects of noise and pointing}
\label{sec:dither_and_noise}

Now we consider noise and telescope pointing together, and we extend the analysis to all four test image configurations.
We again create multiple ``observations'' but now each contains both a different realisation of the noise and a different random pointing.
Table \ref{tab:dither_type} quantifies the spread in best-fit parameter values for all four image configurations and peak S/N values of 1, 10, and 500.

\begin{table*}\centering
\label{tab:dither_type}
\ra{1.3}
\begin{tabular}{@{}rrrrcrrrcrrr@{}}\toprule
& \multicolumn{3}{c}{$S/N = 1$} & \phantom{abc}& \multicolumn{3}{c}{$S/N = 10$} & \phantom{abc} & \multicolumn{3}{c}{$S/N = 500$}\\
\cmidrule{2-4} \cmidrule{6-8} \cmidrule{10-12}
& $lower$ & $median$ & $upper$ && $lower$ & $median$ & $upper$ && $lower$ & $median$ & $upper$\\ \midrule
\textbf{cusp}\\
R${}_E$ & 2.99328 & 2.99564 & 2.99762 && 2.99746 & 2.99919 & 3.00102 && 3.00005 & 3.00014 & 3.00023\\
R.A. & -0.01945 & -0.01232 & -0.00487 && -0.00904 & -0.00372 & -0.00032 && -0.00249 & -0.00177 & -0.00100\\
Dec. & -0.00182 & 0.00290 & 0.00665 && -0.00001 & 0.00291 & 0.00624 && -0.00146 & -0.00084 & -0.00001\\
$e$ & 0.29112 & 0.29323 & 0.29523 && 0.29384 & 0.29620 & 0.29828 && 0.29987 & 0.30019 & 0.30046\\
P.A. & 58.53260 & 58.84365 & 59.13420 && 59.07300 & 59.43300 & 59.71600 && 59.96060 & 59.99050 & 60.02560\\
\\
\textbf{fold}\\
R${}_E$ & 2.99380 & 2.99665 & 3.00037 && 2.99795 & 2.99983 & 3.00218 && 2.99936 & 2.99944 & 2.99957\\
R.A. & -0.00199 & 0.00135 & 0.00526 && -0.00170 & 0.00118 & 0.00322 && -0.00125 & -0.00108 & -0.00093\\
Dec. & -0.01047 & -0.00417 & 0.00234 && -0.00402 & 0.00008 & 0.00250 && -0.00213 & -0.00194 & -0.00168\\
$e$ & 0.29785 & 0.30296 & 0.30593 && 0.29769 & 0.29949 & 0.30265 && 0.30081 & 0.30096 & 0.30108\\
P.A. & 60.20580 & 60.47370 & 60.73450 && 59.93500 & 60.07180 & 60.25280 && 59.99680 & 60.00150 & 60.00790\\
\\
\textbf{cross}\\
R${}_E$ & 2.99555 & 2.99761 & 2.99968 && 2.99678 & 2.99856 & 3.00088 && 2.99965 & 2.99976 & 2.99985\\
R.A. & -0.00175 & 0.00071 & 0.00426 && -0.00266 & 0.00006 & 0.00230 && -0.00009 & 0.00005 & 0.00019\\
Dec. & -0.00738 & -0.00386 & -0.00125 && -0.00347 & -0.00163 & 0.00037 && -0.00002 & 0.00007 & 0.00017\\
$e$ & 0.30288 & 0.30509 & 0.30734 && 0.30090 & 0.30237 & 0.30458 && 0.30031 & 0.30037 & 0.30046\\
P.A. & 59.89760 & 60.18280 & 60.38430 && 59.94060 & 60.05030 & 60.16430 && 59.99650 & 60.00070 & 60.00500\\
\\
\textbf{2-image}\\
R${}_E$ & 2.97860 & 2.99953 & 3.01204 && 2.99677 & 3.00013 & 3.00300 && 2.99906 & 2.99990 & 3.00028\\
R.A. & -0.01245 & -0.00311 & 0.01188 && -0.00307 & 0.00033 & 0.00429 && -0.00198 & -0.00029 & 0.00102\\
Dec. & -0.01217 & -0.00199 & 0.01642 && -0.00356 & -0.00058 & 0.00329 && -0.00276 & -0.00076 & 0.00013\\
$e$ & 0.28867 & 0.30182 & 0.31261 && 0.29752 & 0.29995 & 0.30272 && 0.29957 & 0.30005 & 0.30073\\
P.A. & 59.25010 & 60.40420 & 62.02710 && 59.71070 & 60.01155 & 60.46840 && 59.95650 & 60.00555 & 60.07990\\
\bottomrule
\end{tabular}
\caption{\small
Best-fit lens model parameters when we consider different noise realizations and telescope pointings simultaneously.
The observations correspond to peak S/N values of 1, 10, and 500, and the resolution is 0.05 arcsec/pixel.
For each set of observations, the middle column corresponds to the median value recovered, and the lower and upper bounds of the 68\% CI are shown in the first and third columns, respectively.}
\end{table*}

In general, the scatter decreases as the S/N increases.
The 2-image case tends to have more scatter than the other cases because a 2-image configuration provides weaker constraints than configurations that have additional images and/or long arcs.
The high-S/N cases show some small formal biases in the parameters, but we expect those would be reduced if a PSF were included.

For some applications we are interested in the intrinsic properties of the source galaxy \citep[e.g.,][]{sourceplanescience,sourceplanescience2}.
Depending on the information available, it may be possible to estimate the luminosity, dynamical mass, mass-to-light ratio, gas mass fraction, and star formation rate for the source.
Such applications require knowledge of the lensing magnification, so we examine uncertainties in the magnification associated with noise and pointing.
Specifically, for each lens model in Table \ref{tab:dither_type} we compute the total magnification of the source.
We quantify the scatter using the 68\% confidence interval, and then divide by the true magnification to obtain the fractional uncertainty for each lens configuration.
(We are still just examining the scatter among best-fit models for different realisations of noise and pointing; we are not yet characterising the full uncertainties in individual lens models.)

\begin{figure}
\centering
\includegraphics[width=0.45\textwidth]{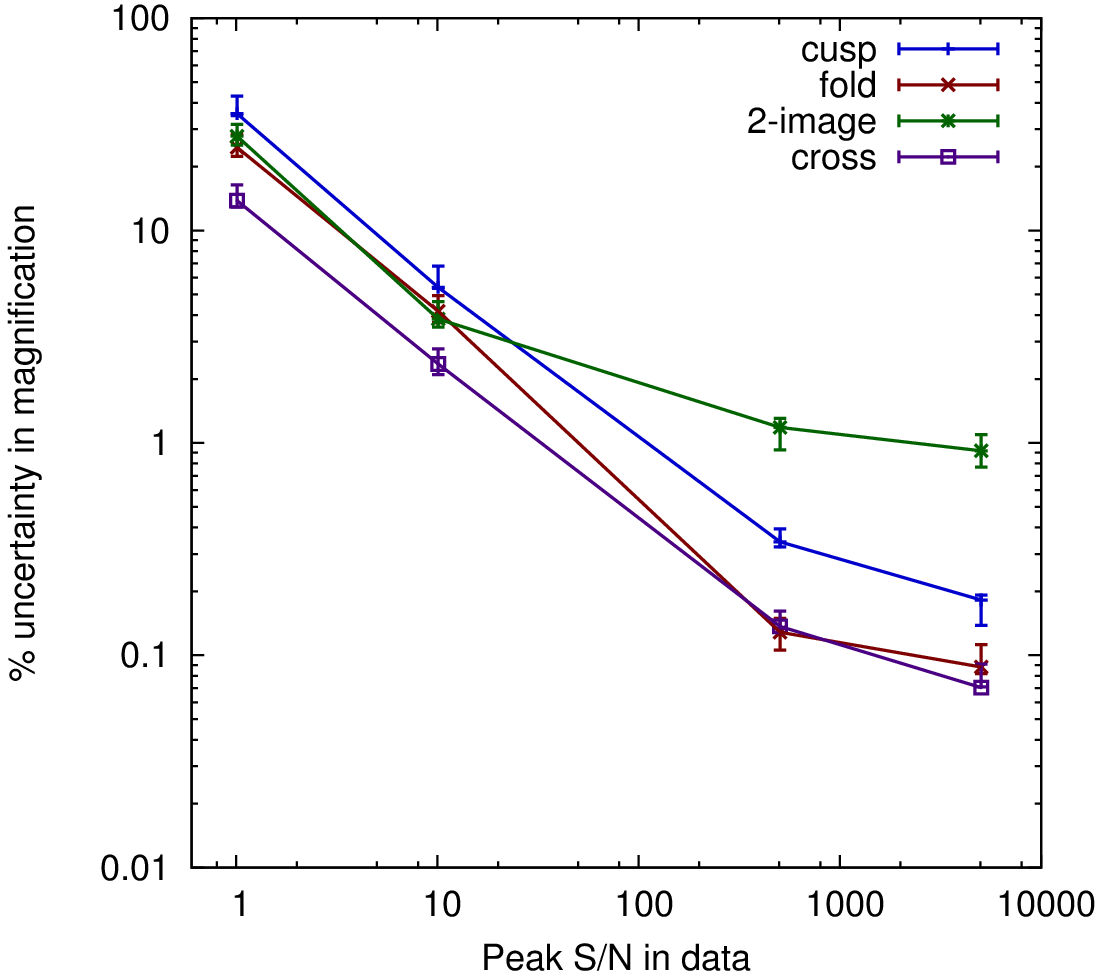}
\caption{\small
Scatter in the lensing magnification (shown as a fractional uncertainty, quoted as a percentage) for different realisations of the noise and telescope pointing.
(This analysis does not take full lens model parameter uncertainties into account.)
Statistical errorbars on the scatter are computed with a bootstrap analysis.
}
\label{fig:mag_fa}
\end{figure}

Fig.~\ref{fig:mag_fa} shows the results.
At low S/N, the cusp configuration has the largest uncertainties, presumably because the source lies in a region where small changes in the model can lead to large changes in the magnification.
At higher S/N, the 2-image case fares worst because the lens model is not highly constrained.
At all S/N values, the cross case has the smallest fractional uncertainties because the source is in a region where the magnification gradient is small.
All told, for $\mathrm{S/N} \gtrsim 10$ the scatter in magnification associated with noise and pointing is $\lesssim 10\%$ for all lens configurations.

%%%%%%%%%%%%%%%%%%%%%%%%%%%%%%%%%%%%%%%%%%%%%%%%%%%%%%%%%%%%
\subsection{Full parameter space exploration}
\label{sec:fullexp}

To this point we have only examined how the best-fit lens model parameters change with different realisations of the noise and telescope pointing.
Now for each ``observation'' we use an adaptive Markov Chain Monte Carlo (MCMC) algorithm to explore the full parameter space and characterise the posterior distribution of parameters.
The width of the posterior depends on the noise level, while the peak location depends on the particular realisation of the noise and pointing.
By comparing the width of each posterior to the scatter across realisations, we can investigate how the scatter from pointing compares to the scatter from noise.
Note that noise contributes to this analysis twice: to the width of each posterior, and to the scatter between them. We consider this ``double counting'' when interpreting the results, as discussed below.

\begin{figure*}
\centering
\includegraphics[width=0.8\textwidth]{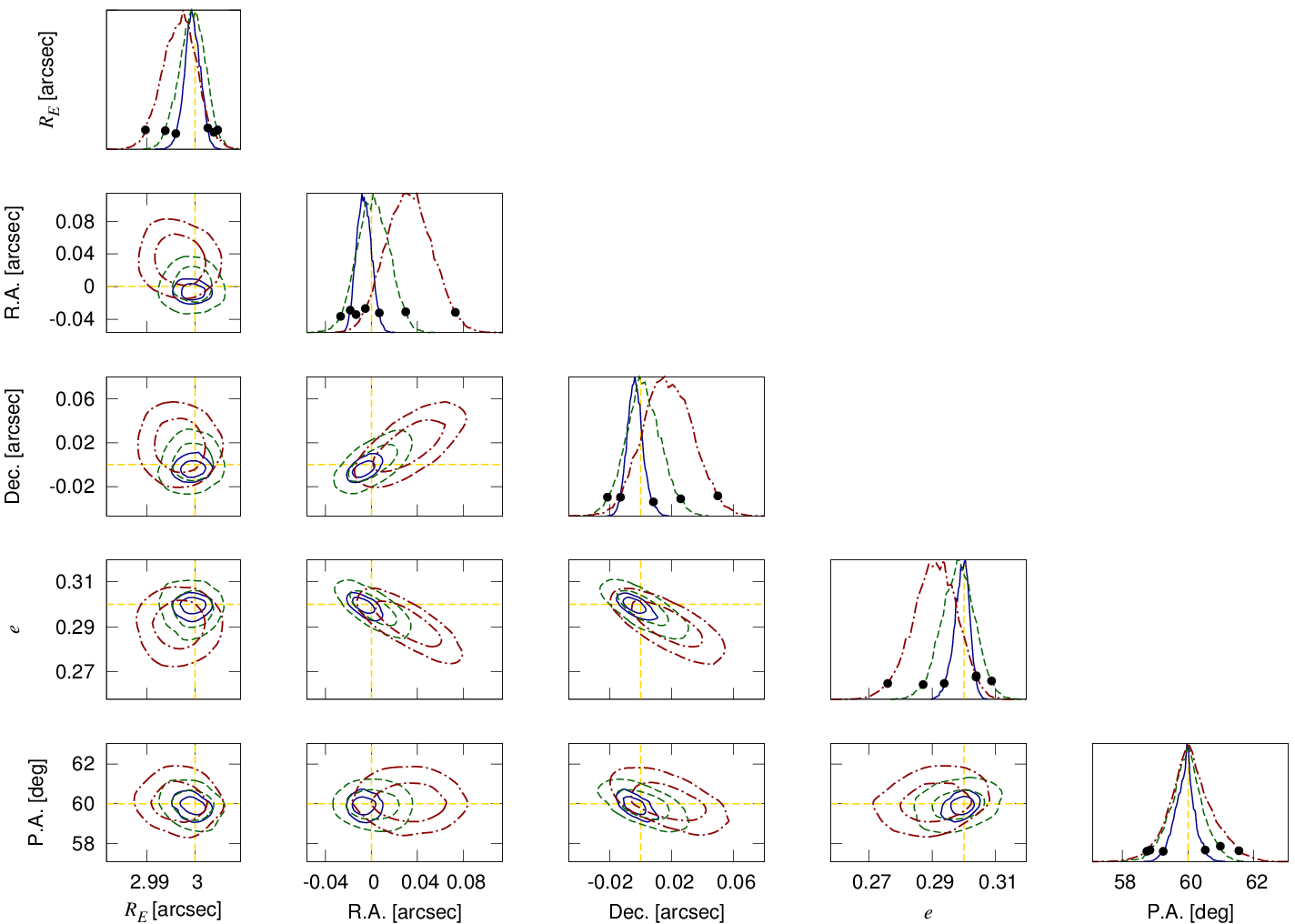}
\caption{\small
Marginal posterior probability distributions for lens model parameters, using a source in the cusp configuration with an image resolution of 0.05 arcsec/pixel and a peak S/N of 10.
We use a Markov Chain Monte Carlo analysis to explore the parameter space and combine 100 realisations of the noise and telescope pointing.
The contour plots show 68\% and 95\% confidence intervals for the various 2-dimensional projections.
The top plots show individual probability distributions (normalised to the same peak), with the 95\% confidence interval marked by points (defined so 2.5\% of the integrated probability is in each of the left and right tails).
Solid blue, dashed green, and dot-dashed red curves correspond to data created with circular Gaussian PSFs having FHWMs of 0.0, 0.12, and 0.24 arcsec, respectively.
}
\label{fig:covariance}
\end{figure*}

Fig.~\ref{fig:covariance} shows the 68\% and 95\% confidence intervals for lens model parameters when we combine all of the realisations.
The noise level is fixed so the peak S/N is 10, and the image resolution is fixed at 0.05 arcsec/pixel.
We analyse the cusp configuration, both without a PSF and with a PSF that has a FWHM of 0.12 or 0.24 arcsec.
Adding a PSF causes the distributions to shift and broaden to some degree, but the true values always lie within the 95\% confidence interval.
Parameter inference, in other words, is robust.

\begin{figure}
  \centering
    \includegraphics[width=0.45\textwidth]{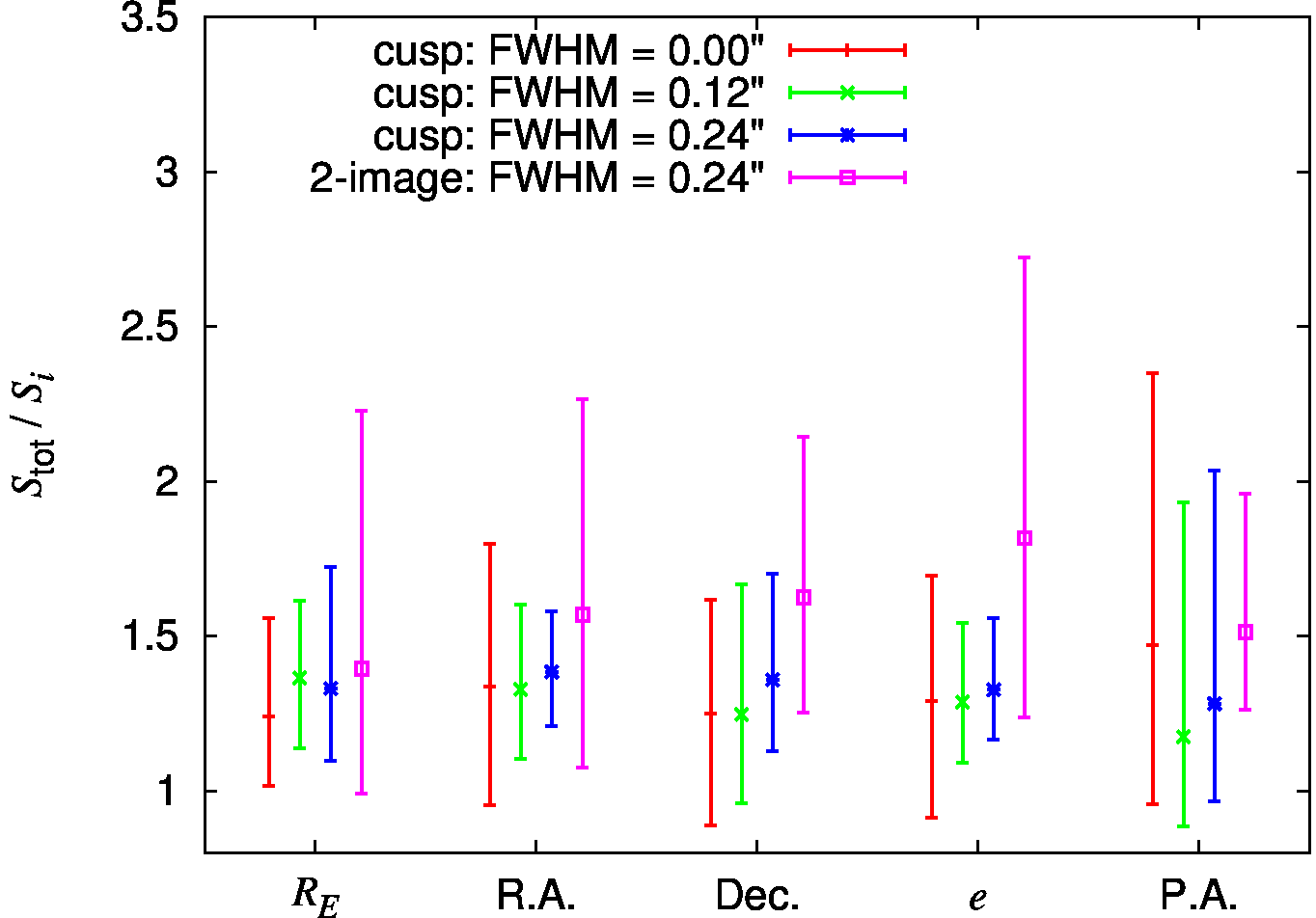} 
  \caption{\small
Ratio of the overall scatter, $S_\text{tot}$, to the scatter for individual runs, $S_i$, for the various lens model parameters.
The point corresponds to the median value of the ratio, while the errorbars are computed with a bootstrap analysis.
A value of $S_\text{tot} / S_i \approx \sqrt{2} \approx 1.4$ indicates that the scatter between runs is comparable to the width of the posterior for an individual run (assuming Gaussianity; see text).
A larger value indicates that there is more scatter between runs.
Results are shown for the cusp configuration with or without a PSF, and the 2-image configuration with a PSF.
The pixel scale is 0.05 arcsec/pixel, and the peak S/N is 10.
}
\label{fig:err_errs}
\end{figure}

Let $S_\text{tot}$ be the width of the posterior from the combined analysis.\footnote{
We quantify the width in terms of the 68\% confidence interval, and use the symbol $S$ to distinguish this scatter from the standard deviation.
}
For comparison, let $S_i$ be the width from an individual ``observation.''
Since $S_i$ only accounts for noise while $S_\text{tot}$ accounts for both noise and pointing, we generally expect $S_\text{tot} > S_i$.
Indeed, Fig.~\ref{fig:err_errs} shows that this ratio typically has values between 1 and 2.
To understand what we might expect, consider that if the distributions were Gaussian then the total scatter would be the quadrature sum of the width of each run and the scatter between runs:
\begin{equation}
  \sigma_\text{tot} \approx \left( \sigma_\text{width}^2 + \sigma_\text{scatter}^2 \right)^{1/2}
  \approx \left( 2\sigma_\text{noise}^2 + \sigma_\text{pointing}^2 \right)^{1/2} ,
\end{equation}
where $\sigma_\text{width} \approx \sigma_\text{noise}$ while $\sigma_\text{scatter} \approx ( \sigma_\text{noise}^2 + \sigma_\text{pointing}^2 )^{1/2}$.
In other words, we might na\"{i}vely predict that the ratio in Fig.~\ref{fig:err_errs} has the form
\begin{equation}
  \frac{S_\text{tot}}{S_i} \approx \left( \frac{2\sigma_\text{noise}^2 + \sigma_\text{pointing}^2}{\sigma_\text{noise}^2} \right)^{1/2} .
\end{equation}
If the scatter from pointing is negligible compared with the scatter from noise, the ratio $S_\text{tot}/S_i$ would have a value near $\sqrt{2} \approx 1.4$.
As the scatter from pointing increases, the ratio would likewise increase.
We could therefore interpret scatter ratios above $\sqrt{2}$ as evidence that scatter due to pointing contributes significantly.

Fig.~\ref{fig:err_errs} does not provide such evidence, however.
The cusp lens configuration scatter ratios that are all consistent with $\sqrt{2}$.
The 2-image configuration has values that are nominally higher but still consistent with $\sqrt{2}$ given the uncertainties.
Therefore, we do not see strong evidence for significant pointing scatter.
While our analytic argument relies on Gaussianity, which may not strictly apply to our distributions, the results suggest that the statsitical properties of our runs are sensible.
We note that these conclusions may depend on the pixel scale and noise level, which we have not explored in detail.

%%%%%%%%%%%%%%%%%%%%%%%%%%%%%%%%%%%%%%%%%%%%%%%%%%%%%%%%%%%%
\section{Conclusions}
\label{sec:conclusions}
%%%%%%%%%%%%%%%%%%%%%%%%%%%%%%%%%%%%%%%%%%%%%%%%%%%%%%%%%%%%

We have introduced a new pixel-based source reconstruction (PBSR) software called \textit{pixsrc} and applied it to mock data in order to investigate statistical and systematic uncertainties in modeling lenses with extended sources.
We have examined several issues that are intrinsic to the pixel-based approach:
\begin{itemize}
\item The $\chi^2$ surface contains ``discreteness noise'' that is influenced by the gridding and regularisation schemes.
\item Errors associated with interpolating surface brightness values in the source plane need to be taken into account, especially for high-S/N data.
\item Adaptive grids are often used to achieve good resolution in the source plane, but they require some care when computing numerical derivatives.
\item A new regularisation scheme called analytic source regularisation (ASR) reconstructs a source with more fidelity than derivative-based regularisation when the data are noisy.
\item Compared to ASR, curvature regularisation may underestimate parameter uncertainties for noisy data.
\end{itemize}
We have applied ASR to sources that are fairly regular, but it could be extended to blended sources or galaxies with star-forming regions by writing the analytic source as a collection of S\'{e}rsic profiles. Differences between ASR and derivative-based regularisation are smaller when the S/N ratio is higher.

We have also examined statistical issues that arise because any given data set has a particular realisation of the noise and telescope pointing.
For the cusp configuration, we find that different realisations of the noise lead to scatter in the best-fit model parameters that scales as a power law in S/N with a slope of $\sim -0.8$.
Different realisations of the pointing lead to scatter that scales as a power law in the pixel scale with a slope of $\sim 3.3$.
Some parameters show small but statistically significant biases, but those can be washed out with the inclusion of a PSF.
When we fully characterise the model uncertainties, the 95\% confidence intervals always include the true parameter values, with or without a PSF.
These results are not highly sensitive to the image configuration, except that our 2-image lens has more scatter than our 4-image lenses because the constraints on the lens model are weaker.

The scatter in noise and pointing lead to scatter in the lensing magnification, which is important for determining the intrinsic properties of the source.
The magnification scatter does vary with the lens configuration because it is sensitive to how rapidly the magnification changes at the location of the source.
This scatter decreases with increasing S/N, but more slowly for the 2-image configuration than for the 4-image cases.
For $\textrm{S/N} \gtrsim 10$ the scatter in magnification associated with noise and pointing is $\lesssim 10\%$ for all lens configurations.

We note that real data may have complications beyond the issues we have addressed.
Examples include irregular structure in the source, differential extinction by dust in the lens galaxy, departures from a smooth lensing potential, incomplete knowledge of the PSF, and intricate aspects of image reduction.
Such issues will be specific to particular data sets and need to be examined in conjunction with the algorithmic issues presented here.

We have discussed a number of different approaches to PBSR, so let us summarise our suggestions for modeling that is both efficient and effective.
If the images can be separated, it may be useful to take their positions and fluxes and perform an initial parameter search assuming point-like images.
Then using an analytic source characterised by a small number of free parameters can help identify the appropriate region of parameter space.
When undertaking full PBSR, derivative-based regularisation is good for computational efficiency, but analytic source regularisation is a valuable step if the source is reasonably well described by an analytic profile or a collection of such profiles.
Finally, it is a good idea to find the best-fit lens and source models and create many realisations of similar observations (as in \S\ref{sec:fullexp}).
That is an effective way to understand the uncertainties and biases in model results given the specific characteristics of the data.

%%%%%%%%%%%%%%%%%%%%%%%%%%%%%%%%%%%%%%%%%%%%%%%%%%%%%%%%%%%%
\section*{Acknowledgements}
%%%%%%%%%%%%%%%%%%%%%%%%%%%%%%%%%%%%%%%%%%%%%%%%%%%%%%%%%%%%

We thank Andrew Baker, Simon Dye, Ross Fadely, Curtis McCully, and Brandon Patel for helpful comments.
We also thank the referee, Olaf Wucknitz, for thorough and helpful comments on the manuscript.
We acknowledge the following computational tools.
For performing Delaunay triangulations, we use the triangle code by \citet{citetriangle}.
For handling world coordinate system information, we use the wcslib library \citep{citewcslib} compiled by WCSTOOLS \citep{citewcstools}.
For solving linear matrix equations and computing matrix factorizations, we use UMFPACK \citep{citeumfpack1,citeumfpack2,citeumfpack3,citeumfpack4}.
For enabling GPU computing, we use CUDA \citep{CUDA} and Thrust \citep{thrust}.
We acknowledge funding from grants AST-0747311 and AST-1211385 from the U.S. National Science Foundation.

\bibliographystyle{mn2e}
\bibliography{refs}

\clearpage

\end{document}